\documentclass[12pt,a4paper]{article}

\usepackage[nosort]{cite}
\usepackage[hyperref,bulletsep]{collect}
\usepackage{graphicx}
\usepackage{pst-all}
\usepackage{bbm}
\usepackage{amsmath}
\usepackage{amssymb}
\usepackage{subfigure}
\usepackage{slashed}
\usepackage{array}

\makeatletter \@addtoreset{equation}{section} \makeatother

\makeatletter
\let\old@startsection=\@startsection
\let\oldl@section=\l@section
\renewcommand{\@startsection}[6]{\old@startsection{#1}{#2}{#3}{#4}{#5}{#6\mathversion{bold}}}
\renewcommand{\l@section}[2]{\oldl@section{\mathversion{bold}#1}{#2}}
\makeatother

\makeatletter
\let\old@makecaption=\@makecaption
\def\@makecaption{\small\old@makecaption}
\makeatother

\renewcommand{\thefootnote}{\arabic{footnote}}
\let\oldPhi=\Phi
\let\oldPsi=\Psi
\let\oldGamma=\Gamma
\let\oldDelta=\Delta
\let\oldSigma=\Sigma
\let\oldTheta=\Theta
\let\oldPi=\Pi
\let\oldUpsilon=\Upsilon
\renewcommand{\Phi}{\mathnormal{\oldPhi}}
\renewcommand{\Psi}{\mathnormal{\oldPsi}}
\renewcommand{\Gamma}{\mathnormal{\oldGamma}}
\renewcommand{\Sigma}{\mathnormal{\oldSigma}}
\renewcommand{\Delta}{\mathnormal{\oldDelta}}
\renewcommand{\Theta}{\mathnormal{\oldTheta}}
\renewcommand{\Pi}{\mathnormal{\oldPi}}
\renewcommand{\Upsilon}{\mathnormal{\oldUpsilon}}

\newcommand{\Action}{\mathcal{S}}

\newcommand{\diag}{\mathop{\mathrm{diag}}}
\newcommand{\sign}{\mathop{\mathrm{sign}}}
\renewcommand{\Re}{\mathop{\mathrm{Re}}}

\newcommand{\order}{\mathcal{O}}

\newcommand{\Integers}{\mathbbm{Z}}
\newcommand{\Reals}{\mathbbm{R}}

\newcommand{\Sphere}{S}  
\newcommand{\AdS}{\mathrm{AdS}}

\ifx\genfrac\sdflkaj

\else

\fi
\newcommand{\sfrac}[2]{{\textstyle\frac{#1}{#2}}}
\newcommand{\half}{\sfrac{1}{2}}

\newcommand{\quarter}{\sfrac{1}{4}}

\newcommand{\matr}[2]{\left(\begin{array}{#1}#2\end{array}\right)}
\newcommand{\alg}[1]{\mathfrak{#1}}
\newcommand{\grp}[1]{\mathrm{#1}}

\newcommand{\grO}{\grp{O}}

\newcommand{\grSO}{\grp{SO}}
\newcommand{\grPSU}{\grp{PSU}}
\newcommand{\algSU}{\alg{su}}

\newcommand{\brk}[1]{(#1)}
\newcommand{\lrbrk}[1]{\left(#1\right)}
\newcommand{\bigbrk}[1]{\bigl(#1\bigr)}
\newcommand{\Bigbrk}[1]{\Bigl(#1\Bigr)}
\newcommand{\biggbrk}[1]{\biggl(#1\biggr)}

\newcommand{\bigsbrk}[1]{\bigl[#1\bigr]}
\newcommand{\Bigsbrk}[1]{\Bigl[#1\Bigr]}
\newcommand{\biggsbrk}[1]{\biggl[#1\biggr]}
\newcommand{\Biggsbrk}[1]{\Biggl[#1\Biggr]}

\newcommand{\lrabs}[1]{\left|#1\right|}
\newcommand{\abs}[1]{{|#1|}}

\newcommand{\Bigabs}[1]{\Bigl|#1\Bigr|}

\newcommand{\eval}[1]{#1|}

\newcommand{\nn}{\nonumber}
\newcommand{\nln}{\nonumber\\}
\newcommand{\nl}[1][0pt]{\nonumber\\[#1]&\hspace{-4\arraycolsep}&\mathord{}}

\newcommand{\earel}[1]{\mathrel{}&\hspace{-2\arraycolsep}#1\hspace{-2\arraycolsep}&\mathrel{}}
\newcommand{\eq}{\earel{=}}

\def\[{\begin{equation}}
\def\]{\end{equation}}

\makeatletter
\def\mr@ignsp#1 {\ifx\:#1\@empty\else #1\expandafter\mr@ignsp\fi}%
\newcommand{\multiref}[1]{\begingroup
\xdef\mr@no@sparg{\expandafter\mr@ignsp#1 \: }%
\def\mr@comma{}%
\@for\mr@refs:=\mr@no@sparg\do{\mr@comma\def\mr@comma{,}\ref{\mr@refs}}%
\endgroup}
\makeatother

\newcommand{\hypref}[2]{\ifx\href\asklfhas #2\else\href{#1}{#2}\fi}

\newcommand{\secref}[1]{Sec.~\multiref{#1}}

\newcommand{\Figref}[1]{Figure~\multiref{#1}}
\newcommand{\figref}[1]{Fig.~\multiref{#1}}
\renewcommand{\eqref}[1]{(\multiref{#1})}

\ifx\href\asklfhas\newcommand{\href}[2]{#2}\fi

\newcommand{\comma}{\quad,\quad}

\newcommand{\tim}[1]{\dot{#1}}

\newcommand{\spa}[1]{\acute{#1}}

\newcommand{\be}{\begin{eqnarray}}
\newcommand{\ee}{\end{eqnarray}}

\newcommand{\Elliptic}[1]{\mathbf{#1}}
\newcommand{\EllipticE}{\Elliptic{E}}
\newcommand{\EllipticF}{\Elliptic{F}}
\newcommand{\EllipticK}{\Elliptic{K}}
\newcommand{\EllipticPi}{\Elliptic{\oldPi}}
\DeclareMathOperator{\am}{am}

\DeclareMathOperator{\sech}{sech}

\DeclareMathOperator{\arcsinh}{arcsinh}

\DeclareMathOperator{\JacobiCN}{cn}
\DeclareMathOperator{\JacobiDN}{dn}
\DeclareMathOperator{\JacobiSN}{sn}

\DeclareMathOperator{\JacobiDC}{dc}
\DeclareMathOperator{\JacobiSC}{sc}
\DeclareMathOperator{\JacobiNC}{nc}

\DeclareMathOperator{\JacobiSD}{sd}

\DeclareMathOperator{\JacobiND}{nd}

\DeclareMathOperator{\JacobiDS}{ds}
\DeclareMathOperator{\JacobiCS}{cs}

\newcommand{\reen}{D} 
\newcommand{\reJ}{\mathcal{J}}
\newcommand{\reE}{\mathcal{E}}

\newcommand{\pws}{p_{\mathrm{ws}}}
\newcommand{\Lelem}{L^{\mathrm{elem}}_{\mathrm{eff}}}
\newcommand{\Ldoub}{L^{\mathrm{doub}}_{\mathrm{eff}}}
\newcommand{\Leff}{L_{\mathrm{eff}}}

\begin{document}

\thispagestyle{empty}
\begin{flushright}\footnotesize
\texttt{arXiv:0803.2324}\\
\texttt{AEI-2008-017}\vspace{0.8cm}
\end{flushright}

\renewcommand{\thefootnote}{\fnsymbol{footnote}}
\setcounter{footnote}{0}

\begin{center}
{\Large\textbf{\mathversion{bold}
Interacting finite-size magnons
}\par}

\vspace{1.5cm}

\textrm{Thomas Klose$^1$ and Tristan McLoughlin$^2$} \vspace{8mm}

\textit{$^{1}$
Princeton Center for Theoretical Physics\\
Princeton University, Princeton, NJ 08544, USA}\\
\texttt{tklose@princeton.edu} \vspace{3mm}

\textit{$^{2}$
Max-Planck-Institut f\"ur Gravitationsphysik\\
Albert-Einstein-Institut\\
Am M\"uhlenberg 1, D-14476 Potsdam, Germany}\\
\texttt{tristan.mcloughlin@aei.mpg.de} \vspace{3mm}


\par\vspace{1cm}

\textbf{Abstract} \vspace{5mm}

\begin{minipage}{14cm}
We explicitly construct a large class of finite-volume two-magnon string solutions moving on $\Reals\times\Sphere^2$. In particular, by making use of the relationship between the $\grO(3)$ sigma model and sine-Gordon theory we are able to find solutions corresponding to the periodic analogues of magnon scattering and breather-like solutions. After semi-classically quantizing these solutions we invert the implicit expressions for the excitation energies in certain limits and find the corrections for the multi-magnon states. For the breather-like solutions we express the energies directly in terms of the action variable whereas for the scattering solution we express the result as a combination of corrections to the dispersion relation and to the scattering phase.
\end{minipage}

\end{center}

\vspace{0.5cm}

\newpage
\setcounter{page}{1}
\renewcommand{\thefootnote}{\arabic{footnote}}
\setcounter{footnote}{0}

\tableofcontents

\newpage
\section{Introduction}

The solution to the problem of finding the spectrum of the ${\cal N}=4$ super-Yang-Mills dilatation operator or equivalently, by the conjectured AdS/CFT duality \cite{Maldacena:1997re}\cite{Witten:1998qj,Gubser:1998bc}, that of finding the energies  of quantum strings on $AdS_5\times S^5$ has seen significant progress in recent times. In particular the discovery that the perturbative dilatation operator in the planar limit is described by an integrable spin-chain Hamiltonian \cite{Minahan:2002ve,Beisert:2003yb,Beisert:2003tq} and of the existence of classical integrability for the string sigma model \cite{Mandal:2002fs,Bena:2003wd} has lead to the introduction of a range of new powerful tools. As is the case for many integrable models the dispersion relation for the fundamental excitations and the two body S-matrix provide a complete description of the theory in infinite volumes. That the fundamental excitation, the  magnon, dispersion relation is given by the BMN result
\cite{Berenstein:2002jq}
\be
\Delta=\sqrt{1+\frac{\lambda}{\pi}\sin^2 \frac{p}{2}}
\ee
was shown to follow from the existence of global symmetries preserved by the spin chain vacuum state \cite{Beisert:2005tm}. The relevant S-matrix for studying the infinite volume spectrum was introduced by \cite{Staudacher:2004tk} and,  remarkably, was  fixed  up to an overall undetermined phase by the global symmetries in \cite{Beisert:2005tm}.  Furthermore an asymptotic strong coupling expansion for the phase itself was conjectured by Beisert, Hernandez and Lopez in  \cite{Beisert:2006ib} based on  perturbative results \cite{Arutyunov:2004vx, Hernandez:2006tk,Gromov:2007cd} and compatibility with the crossing symmetry as formulated by \cite{Janik:2006dc}. This conjecture was subsequently extended by Beisert, Eden and Staudacher \cite{Beisert:2006ez}, via an inspired ``analytic continuation'', to a weak coupling expansion which begins at fourth order in a loop expansion and which has passed several checks against perturbative calculations \cite{Bern:2006ew, Cachazo:2006az, Beisert:2007hz}. This S-matrix provides, via the asymptotic Bethe ansatz \cite{Beisert:2005fw}, predictions for the all-order anomalous dimensions of operators with infinitely large charges. 

Nonetheless there remain several outstanding issues amongst which is the pressing question of what happens for operators of finite charge. There is strong evidence that the long range interactions of the higher loop terms in the spin chain Hamiltonian give rise to so called wrapping-interactions which spoil the application of the Bethe ansatz. General considerations using the thermodynamic Bethe ansatz \cite{Ambjorn:2005wa} show that the wrapping effects will generically occur at the $L$-th loop order for spin chains of length $L$ and more concretely it was shown \cite{Kotikov:2007cy} that the anomalous dimensions for finite-spin twist-two operators predicted by asymptotic Bethe ansatz disagreed with constraints from the BFKL behavior of high energy scattering amplitudes. Relatedly, direct perturbative calculations of the four-loop anomalous dimension of the length four Konishi operator have shown the presence of wrapping effects \cite{Keeler:2008ce,Fiamberti:2007rj}, see also \cite{Sieg:2005kd,Eden:2007rd}. (It should be mentioned that there is currently a discrepancy between these independent calculations.)

The string duals to operators with infinite charges are quantum states of the world-sheet theory defined on the infinite plane. This is perhaps seen most clearly in the physical light-cone gauge where the theory is defined on a cylinder with radius proportional to the light-cone momentum, a combination of the $\AdS$ energy and an angular momentum from the compact space. Taking the infinite angular momentum limit corresponds to decompactifying the cylinder and one can now consistently define a S-matrix for the resulting massive, non-Lorentz invariant, integrable world-sheet theory. This world-sheet S-matrix can be calculated perturbatively and has been show to reproduce the tensor structure \cite{Klose:2006zd} of the exact conjectured S-matrix at leading order and, in the near-flat limit \cite{Maldacena:2006rv}, reproduce the phase to two-loops \cite{Klose:2007wq,Klose:2007rz}. Furthermore it has been shown that the conjectured S-matrix is consistent with the Zamolodchikov-Faddeev algebra following from the conjectured integrability of the string sigma-model \cite{Arutyunov:2006yd}. One particularly elegant result by Hofman and Maldacena \cite{Hofman:2006xt} is the construction of the string dual to the elementary spin chain excitation, the giant magnon, which has a classical dispersion relation
\be
\Delta\simeq \frac{\sqrt{\lambda}}{\pi}\left|\sin\frac{p}{2}\right|.
\ee
These giant magnons are rigid open strings moving on $\Reals\times\Sphere^2$  with infinite angular momentum; they have an infinitely extended world sheet and the magnon momentum corresponds to the opening angle of the string end points viewed from the center of $\Sphere^2$. By considering multi-magnon solutions it was further possible to calculate  the semi-classical scattering phase and show that it agreed with the previously calculated semi-classical S-matrix of Arutyunov, Frolov and Staudacher (AFS). 

Moving from the theory defined on the plane to the finite volume theory presents significant challenges and to date there have been only limited results. Explicit calculations, \cite{Schafer-Nameki:2005tn}, of  quantum corrections to the energies of rigid spinning strings were shown to disagree with the predictions of from the Bethe ansatz with corrections that are exponentially small in the string length \cite{Schafer-Nameki:2006ey}. An alternative approach has been to consider the finite volume corrections coming from the finite angular momentum, $J$, analogues of the giant magnons. The dispersion relation for such finite angular momentum solutions was calculated in \cite{Arutyunov:2006gs} for a variety of gauge choices. The resulting corrections to the single magnon dispersion relation were exponentially suppressed in the effective string length and gauge dependent. The gauge dependence followed from the fact that in finite volume it was not possible to construct a consistent string with non-vanishing magnon momentum though this obstacle could be overcome by considering the string moving on a orbifold of the $S^5$ \cite{Astolfi:2007uz}. Just as the giant magnon, a string moving on an $\Sphere^2\subset\Sphere^5$, could be generalized to dyonic bound states \cite{Dorey:2006dq}, strings in $\Sphere^3$, their finite size counterparts can be similarly generalized \cite{Okamura:2006zv} and the corrections to the dispersion relation calculated \cite{Hatsuda:2008gd}. Another method  makes use of arguments of L\"{u}scher \cite{Luscher:1985dn}, generalized to the non-Lorentz invariant world-sheet theory  \cite{Janik:2007wt,Heller:2008at}, to use the asymptotic S-matrix to calculate the leading order exponential corrections to the dispersion relation. Indeed this method can be extended to calculate not merely classical corrections but corrections coming from quantum fluctuations \cite{Gromov:2008ie}. A very general approach to studying integrable theories in finite volume is the thermodynamic Bethe ansatz, unfortunately this usually makes use of the Lorentz invariance of the two-dimensional theory; generalizing the relevant results to the world-sheet theory is a significant challenge though some progress has already been made \cite{Arutyunov:2007tc}.

It is obviously of interest to consider the finite size corrections to multi-magnon states: for one thing it makes it possible to form physical closed strings but perhaps more importantly it allows calculation of the finite volume effects on the magnon interactions. While the concept of asymptotic states no longer makes sense on a cylinder and so it is not possible to define an S-matrix it is of course possible to calculate the corrections to the energies of multi-magnon states. As it has been the case for the infinite volume theory it may be that understanding perturbative results on the world sheet can provide hints for the exact answer and will certainly provide checks of whatever conjectures are made. Already studies of multi giant magnon states at  leading order in exponential corrections have been carried out using algebraic curve methods \cite{Minahan:2008re}. For a more concrete description we would like to find string solutions corresponding to these multi-magnon states. Discouragingly the string equations of motion are non-linear and while, by using standard finite-gap methods, it is possible to find a general abstract solution \cite{Dorey:2006zj} it is difficult to find explicit solutions that are simple enough to manipulate. Instead it is possible to make use of the relation between strings on ${\mathbb R}\times S^2$ and sine-Gordon theory which was first described in \cite{Pohlmeyer:1975nb} and discussed in the context of AdS/CFT by \cite{Mikhailov:2005qv,Mikhailov:2007xr,Grigoriev:2007bu}. This purely classical correspondence relates the string equations of motion and constraints to the sine-Gordon equation of motion: in the simplest case the giant magnon corresponds to the kink solution of sine-Gordon theory. Similarly, it was by using the correspondence between two magnon states and kink anti-kink states that \cite{Hofman:2006xt} was able to calculate the semi-classical scattering phase and breather spectrum. For finite volume the counting of distinct excitations is not so reliable and instead we classify the different solutions by the number of independent arguments, called phases, on which they depend. For example the finite-$J$ one magnon solution of \cite{Arutyunov:2006gs} corresponds to the single phase kink-train of sine-Gordon theory, e.g. \cite{Forest:1982ji}. In this work we make use of the known two-phase solutions of sine-Gordon theory \cite{Costabile:1978}. These were constructed using the Lamb ansatz \cite{Lamb:1971zz} where the sine-Gordon field, $\phi(x,t)$, is assumed to be of the form
\be
  \phi(x,t)=2 \arctan F(t)G(x) \; ,
\ee
with $F$ and $G$ even functions satisfying ordinary differential equations which can be solved in terms of Jacobi elliptic functions. This may seem surprising as the generic two-phase solution leads to hyperelliptic functions on a genus two Riemann surface. However, if the initial conditions are symmetric about $x=0$ then the solutions are standing waves whose $x$ and $t$ flows separate and can be expressed in terms of elliptic functions \cite{Forest:1982ji}.

Reconstructing the target-space string from a given sine-Gordon solution is in general a very non-trivial problem, fortunately it is possible to make use of the classical relations between sine-Gordon theory and the geometry of constant curved surfaces, see e.g. \cite{Gu:1996zm}. We explicitly integrate the equations describing the string surfaces, find their embedding for all values of their parameters, and calculate their global charges. We find a rich moduli space of solitons consisting of the periodic analogues of magnon scattering solutions and magnon `breathers'%
\footnote{We call the reconstruction of breather-like sine-Gordon solutions magnon breathers. As discussed in \cite{Dorey:2007xn} in the decompactification limit they are superpositions of BPS magnons carrying opposite charges and with both magnons having real kinematic variables. In order to see this one must go to the larger space $\Reals\times\Sphere^3$ where it can be shown that they arise from the two-charge magnon solutions of Spradlin and Volovich \cite{Spradlin:2006wk}. Not having the periodic solutions for this larger sector we are not able to unambiguously determine whether the same interpretation persists for the string solutions corresponding to both the fluxon and plasmon breather though as they both reduce to the same breather in the decompactification limit this seems likely.}. %
All of these solutions are periodic in time as well as in the spatial coordinate and so we can use Bohr-Sommerfeld quantization to express their energies in terms of a single integer quantum number, $n$, which for the fluxon oscillation solutions is related to the magnon momentum and for breathers to the usual action variable. 

An outline of the paper is as follows: in \secref{sec:String_action} we briefly describe the string action and its relation via Pohlmeyer reduction to sine-Gordon. In \secref{sec:sg_solitons} we recall some of the known solutions of sine-Gordon theory including the periodic two-phase solutions: the fluxon oscillation, the breather and the plasmon. We perform the reconstruction, solve the inverse map and find the explicit periodic solutions to the string equations of motion in \secref{sec:reconstruction}. In addition we find explicit formulae for the angular momenta of the individual solutions. In  \secref{sec:quantization} after semi-classically quantizing the solutions we find expressions for the energy formulae of the magnon breathers by expanding in the near-decompactification limit where we find the finite size corrections the breather solutions found by Hofman and Maldacena. In addition we are able to compare our solution for the case of $J=0$ with the pulsating circular string found by Minahan \cite{Minahan:2002rc} where we find agreement. Furthermore using the relation between the momentum, phase-shift and oscillation number we identify the dispersion relation for the single and double magnon solutions. For the appropriate solutions, and in the appropriate limits,  we match those corrections with those previously found in the literature. Additionally we find the corrections to the periodic analogue of the scattering phase in the center of mass frame.

\section{From strings on $\Reals\times\Sphere^2$ to sine-Gordon}
\label{sec:String_action}

Superstrings living in a $AdS_5\times S^5 $ background can be described by the Green-Schwarz-Metsaev-Tseytlin action for the supercoset $\tfrac{\grPSU(2,2|4)}{\grSO(1,4)\times\grSO(5)}$ \cite{Metsaev:1998it}. We will focus on closed bosonic strings moving in an ${\mathbb R}\times S^2$ subspace, which as we are only interested in classical solutions, is a consistent truncation. We fix part of the world-sheet diffeomorphism invariance by choosing the world-sheet metric to be conformally flat  $h^{ab} \propto \diag(+,-)$ (conformal gauge) so the string action becomes,
\be
  \Action = \frac{\sqrt{\lambda}}{4\pi} \int_{-\infty}^\infty\!d\tau \int_0^L\!d\sigma\: \biggsbrk{ - (\partial t)^2 + (\partial\vec{n}) + \Lambda (\vec{n}^2 - 1) }.
\ee
The constrained three-vector, $\vec{n}=(\cos \phi \sin \theta, \sin \phi \sin \theta, \cos \theta)$, describes the string on a $\Sphere^2$ with unit radius as the overall size, $R$, has been absorbed into the string tension to form the 't Hooft coupling, $\lambda$. The world-sheet is a cylinder with circumference $L$ and we fix the remaining gauge-freedom by identifying $\tau$ with the target-space time, $t =  \tau$, after which the string is simply described by an $\grO(3)$ sigma model. Denoting derivatives by subscripts, $\vec{n}_\tau \equiv \partial_\tau \vec{n}$ etc, the equations of motion (after solving for the Lagrange multiplier) and the Virasoro constraints are 
\be 
\label{eqn:eom_tau_sigma}
  & &\vec{n}_{\tau\tau} - \vec{n}_{\sigma\sigma} = - \bigsbrk{(\vec{n}_\tau)^2 - (\vec{n}_\sigma)^2} \vec{n} \; , \\
\label{eqn:virasoro_tau_sigma}
& &\kern-10pt  (\vec{n}_\tau)^2 + (\vec{n}_\sigma)^2 = 1
	\comma
  \vec{n}_\tau \cdot \vec{n}_\sigma = 0	\; .
\ee
It is worth noting that in this case the constraints are simply that world-sheet energy density is a constant and that momentum density is zero. It can also easily seen that the equations of motion follow directly from the constraint equations and, conversely, any solution to the equation of motion automatically satisfies the constraints. 

For strings moving on $\Reals\times\Sphere^2$ the relevant global charges are the target-space energy,  
\be \label{eqn:target-space-energy}
E = \frac{\sqrt{\lambda}}{2\pi} \, \reE
  \comma
  \reE = \int_0^L\!d\sigma\: t_\tau = L \; ,
\ee
which in the static gauge is just the string length $L$, and the angular momentum
\be
  \vec{J} = \frac{\sqrt{\lambda}}{2\pi} \, \vec{\reJ}
  \comma
  \vec{\reJ} = \int_0^L\!d\sigma\: \vec{n}\times \vec{n}_\tau \; .
\ee
The closed string states will be classified by the $\algSU(2)$ Cartan element, $J_3$, and the mass shell condition (conformal constraint) will give the target-space energy as a function of the coupling, the  $\algSU(2)$ charge, and, after quantization, the relevant quantum  numbers. By the conjectured duality this energy should be equivalent to the scaling dimension of a single trace operators with the same quantum numbers.

In this work we wish to make use of the classical equivalence, first derived by Pohlmeyer \cite{Pohlmeyer:1975nb}, between the $\grO(3)$ sigma model and sine-Gordon theory. To describe this relation it is useful to introduce light-cone coordinates, $\sigma^\pm = \tau \pm \sigma$, $\partial_{\pm} = \half\brk{\partial_\tau \pm \partial_\sigma}$, so that the equation of motion and the constraints become
\be
& & \vec{n}_{+-} = - (\vec{n}_+\cdot\vec{n}_-) \vec{n} \\
& &
\label{eqn:constraints_lightcone}  
  \vec{n}^2 = 1
  \comma
  \vec{n}_+^2 = \tfrac{1}{4}
  \comma
  \vec{n}_-^2 = \tfrac{1}{4} \; .
\ee
Given the equivalence of the constraints to the equations of motion solving the system corresponds to finding three vectors, $\vec{n}$, $\vec{n}_+$, $\vec{n}_-$, satisfying \eqref{eqn:constraints_lightcone}. The solution is not unique and different solutions are distinguished by the angle between $\vec{n}_+$ and $\vec{n}_-$ as a function of $\sigma^\pm$. Defining $\phi$ to be half of this angle,
\be \label{eqn:defSGfield}
  \cos 2\phi = 4 \vec{n}_+\cdot\vec{n}_- = (\vec{n}_\tau)^2 - (\vec{n}_\sigma)^2 \; ,
\ee
one can show \cite{Pohlmeyer:1975nb} that all conditions on $\vec{n}$ are equivalent to the sine-Gordon equation of motion for $\phi$, i.e.
\be \label{eqn:SGeq}
  \phi_{+-} = - \frac{1}{8} \sin2\phi
  \qquad\mbox{or}\qquad
  \phi_{\tau\tau} - \phi_{\sigma\sigma} = - \frac{1}{2} \sin2\phi
  \; .
\ee
Thus to every solution of the sine-Gordon equation there is a solution to the string equations of motion satisfying the constraints. The string boundary conditions, however, have to be imposed additionally. Moreover, the presence of the derivatives in \eqref{eqn:defSGfield} makes the inverse map non-local and therefore complicated. It is also important to stress that the equivalence is only at the level of the classical equations of motion; in fact the two theories have different Poisson structures and certainly the two theories are quite different when considered quantum mechanically. The Pohlmeyer reduction or reformulation has been generalized to other systems; for example  the  $\grO(4)$ sigma model, which is equivalent to complex sine-Gordon \cite{Pohlmeyer:1975nb}, \cite{Lund:1976ze,Lund:1976xd}, and interestingly the full superstring on the supercoset $\tfrac{\grPSU(2,2|4)}{\grSO(1,4)\times\grSO(5)}$ \cite{Mikhailov:2007xr,Grigoriev:2007bu}. Because of the special properties of the superstring in the $\AdS_5\times \Sphere^5$, for one the theory is conformal even quantum mechanically, it has been speculated that the equivalence in this case may possibly extend to the quantum theory.

\section{Soliton solutions of sine-Gordon theory}
\label{sec:sg_solitons}

In this section we describe the known periodic two-phase solutions of the sine-Gordon equation \eqref{eqn:SGeq}. These are the solutions that represent two interacting giant magnons with finite angular momentum; see \secref{sec:reconstruction} for the reconstruction of the classical string from these solutions. 

In order to study closed strings with finite world-sheet circumference $L$, we need to impose (quasi-)periodic boundary conditions
\be
  \phi(\sigma+L,\tau) = \phi(\sigma,\tau) \;\; \mod 2\pi
\ee
on the sine-Gordon field $\phi(\sigma,\tau)$. The classification of solitonic solutions is most conveniently done by counting the number of independent linear combinations of $\sigma$ and $\tau$ on which the solution depends. These combinations are referred to as phases. In infinite volume, counting phases is equivalent to counting the number of kinks, but in the case of some periodic solutions counting kinks is misleading and does not lead to a sensible classification.

\begin{figure}[!ht]
\begin{center}
\subfigure[Single kink]{\includegraphics[width=44mm]{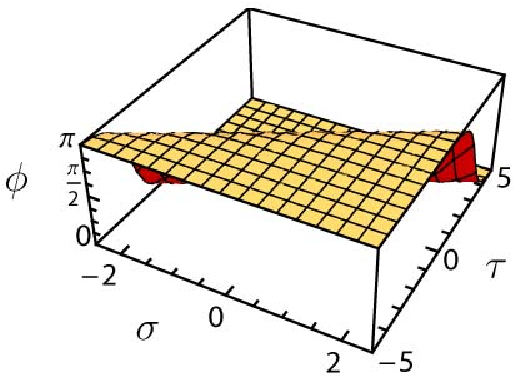}}\hspace{5mm}
\subfigure[Kink--anti-kink scattering]{\includegraphics[width=44mm]{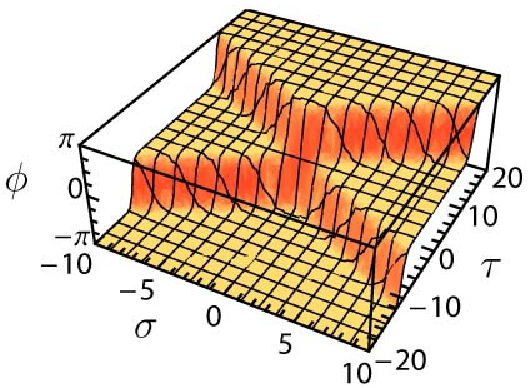}} \hspace{5mm}
\subfigure[Breather]{\includegraphics[width=44mm]{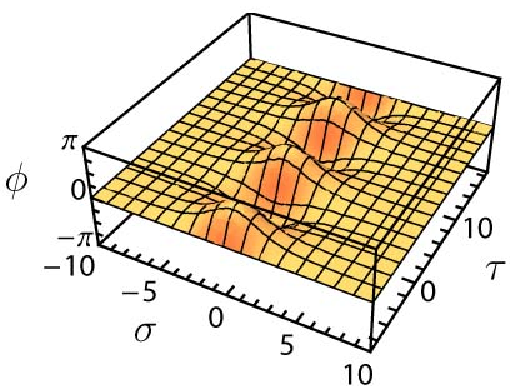}} \\
\subfigure[Kink train (elem)]{\includegraphics[width=44mm]{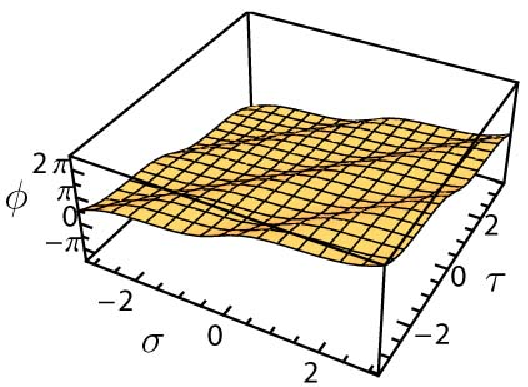}
  \label{fig:sol-single-finite-e}} \hspace{5mm}
\subfigure[Fluxon oscillation (elem)]{\includegraphics[width=44mm]{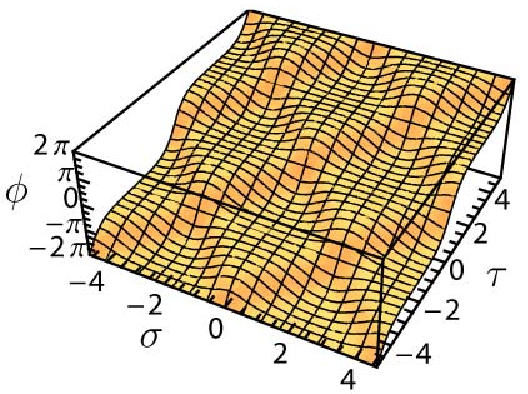}
  \label{fig:sol-scatt-finite-e}} \hspace{5mm}
\subfigure[Fluxon breather (elem)]{\includegraphics[width=44mm]{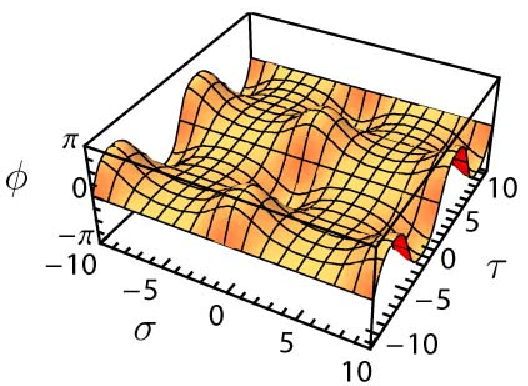}
  \label{fig:sol-bound-finite-e}} \\
\subfigure[Kink train (doub)]{\includegraphics[width=44mm]{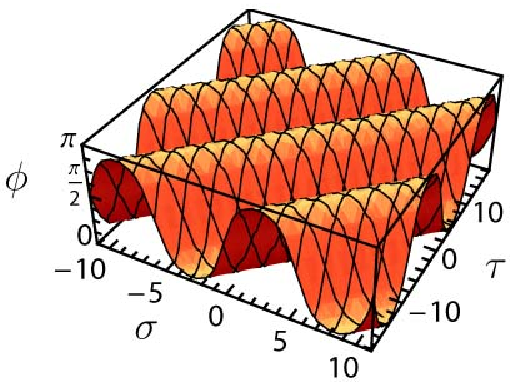}
  \label{fig:sol-single-finite-d}} \hspace{5mm}
\subfigure[Fluxon oscillation (doub)]{\includegraphics[width=44mm]{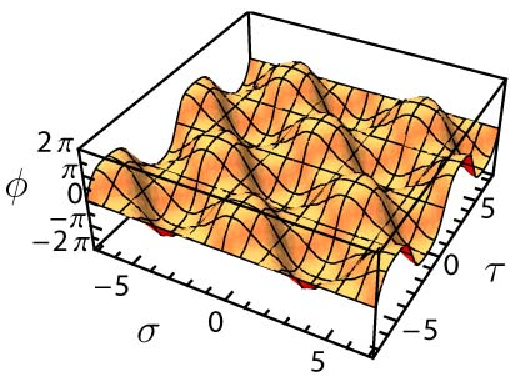}
  \label{fig:sol-scatt-finite-d}} \hspace{5mm}
\subfigure[Fluxon breather (doub)]{\includegraphics[width=44mm]{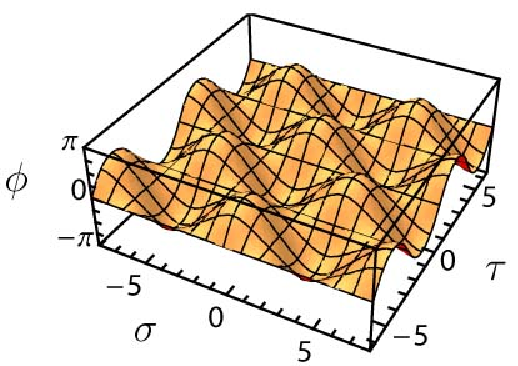}
  \label{fig:sol-bound-finite-d}}
\caption{\textbf{Sine-Gordon solitons.} The first row shows the basic one and two soliton configurations on the infinite line. Their periodic generalizations are shown below, in the elementary (second row) and the doubled region (third row). The plasmon breather looks qualitatively the same as the fluxon breather and has the same decompactification limit (c). The periodic solutions are plotted over two periods, i.e. $2L$ and $2T$. Note that in the cases (d) and (e) we identify $\phi \sim \phi + 2\pi$ to make the solution strictly periodic.}
\label{fig:solitons}
\end{center}
\end{figure}

In fact, the general $n$-phase solution can be written down in terms of Riemann theta functions \cite{Matveev:1976mj,Kozel:1976}.
Here we are only interested in the special but broad class of the two-phase solutions which can be written in Lamb form \cite{Lamb:1971zz}
\be \label{eqn:lamb-form}
  \phi(\sigma,\tau) = 2 \arctan F(\tau) G(\sigma) \; .
\ee
They were first studied in \cite{Costabile:1978} as a description of the magnetic flux in a Josephson junction between superconductors. All quasi-periodic two-phase solutions of the form \eqref{eqn:lamb-form} can be divided into three types known as the fluxon oscillation, the fluxon breather and the plasmon breather%
\footnote{In solid state physics the usual nomenclature is `fluxon' for the fluxon oscillation, `breather' for the fluxon breather and `plasmon' for the plasmon breather \cite{Fulton:1977}.} %
\cite{Costabile:1978}. They correspond to different combinations of Jacobi elliptic functions for $F(\tau)$ and $G(\sigma)$ and will be reviewed in the following. We also include their decompactification limits, $L\to\infty$, as well as the one-phase solutions into this overview. \Figref{fig:solitons} displays the relevant soliton solutions.

\subsection{Single kink}

The fundamental soliton solutions on the infinite line are the single kink ($+$) and the single anti-kink ($-$) given by
\be \label{eqn:SG-single-decomp}
  \phi(\sigma,\tau) = 2 \arctan e^{\pm\gamma (\sigma-\beta\tau)}
\ee
with $\gamma=1/\sqrt{1-\beta^2}$. They are one-phase solutions as they depend on $\sigma$ and $\tau$ only through the linear combination $\sigma-\beta\tau$. The free parameter $\abs{\beta}<1$ represents the velocity of the soliton. Notice also the Lamb form \eqref{eqn:lamb-form} of this solution.

The string corresponding to this sine-Gordon field is the Hofman-Maldacena giant magnon \cite{Hofman:2006xt}. Although it lives on a decompactified world sheet, the length of this string in target space is finite.

\subsection{Kink scattering}

The solution describing the scattering of a kink and an anti-kink is given by
\be \label{eqn:SG-scatt-decomp}
  \phi(\sigma,\tau) = 2 \arctan \frac{\sinh \gamma \beta \tau}{\beta \cosh \gamma \sigma}
\ee
with $\gamma$ as above. This is a two-phase solution of Lamb form where the parameter $\beta$ denotes the relative velocity in the center of mass frame. From this solution, one can obtain the scattering of two kinks by the shift
\be \label{eqn:shift-asol-to-sol}
  \gamma\sigma \to \gamma\sigma + \frac{i\pi}{2}
  \comma
  \gamma\beta\tau \to \gamma\beta\tau + \frac{i\pi}{2} \; ,
\ee
which results in
\be \label{eqn:SG-ss-scatt-decomp}
  \phi(\sigma,\tau) = 2 \arctan \frac{\cosh \gamma \beta \tau}{\beta \sinh \gamma \sigma} \; .
\ee
These solutions correspond to the scattering of two giant magnons \cite{Hofman:2006xt}.

\subsection{Breather}

By analytically continuing the velocity in the scattering solution \eqref{eqn:SG-scatt-decomp} to $\beta \to i a$, one obtains the breather
\be \label{eqn:SG-bound-decomp}
  \phi(\sigma,\tau) = 2 \arctan \frac{\sin\gamma_a a \tau}{a \cosh\gamma_a \sigma}
\ee
with $\gamma_a=1/\sqrt{1+a^2}$. This solution is periodic in $\tau$ with period $T=\frac{2\pi}{a\gamma_a}$ and describes a bound state of a kink and an anti-kink. A bound state of two kinks does not exist; the analytic continuation of \eqref{eqn:SG-ss-scatt-decomp} would produce an imaginary sine-Gordon field.

The solution \eqref{eqn:SG-bound-decomp} corresponds to a superposition of two giant magnons with opposite charges \cite{Dorey:2007xn}.

\subsection{Kink train}

Now, we turn to $\sigma$-periodic sine-Gordon fields which, as we will see later, give rise to strings with finite world sheet. The fundamental periodic soliton solutions are given by the kink train ($+$) and the anti-kink train ($-$)
\be \label{eqn:SG-single-finite}
  \phi(\sigma,\tau) = \frac{\pi}{2} + \am\bigbrk{\pm(k\sigma-\omega\tau)|m} \; .
\ee
These solutions contain two independent parameters $k$ and $\omega$ which determine the elliptic modulus as
\be
  m = \frac{1}{k^2-\omega^2} \; .
\ee
For $m<1$, which implies $k^2<\omega^2$ or $k^2>\omega^2+1$, the solution describes an infinite sequence of kinks (or anti-kinks) moving with fixed velocity $\omega/k$ and equal separation given by the spatial period%
\footnote{This formula also hold in the cases $k^2=\omega^2$ and $k^2=\omega^2+1$ where $m=\pm\infty$ and $m=1$, respectively.} %
\be \label{eqn:L-kink-train-elem}
  L = \frac{2}{k}  \, \EllipticK(m) \qquad\mbox{for $k^2\le\omega^2$ or $k^2\ge\omega^2+1$} \; .
\ee
As every kink is a step of $2\pi$, this solution is only \emph{quasi}-periodic, see \figref{fig:sol-single-finite-e}. Since every interval of length $L$ contains exactly one soliton, we call this region of parameter space the `elementary region'.

For $m>1$, i.e. $\omega^2<k^2<\omega^2+1$, there is an anti-kink (kink) inserted between any two kinks (anti-kinks) of the infinite sequence moving with the same velocity, see \figref{fig:sol-single-finite-d}. These insertions make the field strictly periodic with period
\be \label{eqn:L-kink-train-doub}
  L = \frac{4}{k\sqrt{m}}  \, \EllipticK\bigbrk{\tfrac{1}{m}} \qquad\mbox{for $\omega^2<k^2<\omega^2+1$} \; .
\ee
Because the insertions do not occur precisely in the middle of two kinks (anti-kinks), the two cases are not related by a shift but by a reflection in $\sigma$ and $\tau$. Since every period contains one kink and one anti-kink, we call this region of parameter space the `doubled region'.

For $m=1$, or $k^2 = \omega^2+1$, the periods become infinite. Thus, sending $m\to1$ is the decompactification limit and the solution from both regions go smoothly into \eqref{eqn:SG-single-decomp} with the identification $k = \gamma$ and $\omega = \gamma\beta$. For the solution in the doubled region the kinks of the opposite kind are pushed infinitely far from  the, itself infinite, region captured by the elementary decompactified solution. In this respect we note the factor of two difference in the prefactor of the periods \eqref{eqn:L-kink-train-elem} and \eqref{eqn:L-kink-train-doub}. In a sense this makes the decompactified doubled solution twice infinite and gives enough room to include the mirror kink.

The final case $k^2=\omega^2$ should be excluded as there $m$ diverges and the solution becomes arbitrarily oscillatory.

Being a one-phase solution, the periodicity in $\sigma$ implies a periodicity in $\tau$. The temporal (quasi-)periods are given by $\frac{k}{\omega} L$, i.e.
\be \label{eqn:T-kink-train}
  T = \begin{cases}
      \frac{2}{\omega}  \, \EllipticK(m) & \mbox{for $k^2\le\omega^2$ or $k^2\ge\omega^2+1$} \; , \\
      \frac{4}{\omega\sqrt{m}}  \, \EllipticK\bigbrk{\tfrac{1}{m}} & \mbox{for $\omega^2<k^2<\omega^2+1$} \; .
      \end{cases}
\ee

We note that the solution \eqref{eqn:SG-single-finite} is not of factorized Lamb form \eqref{eqn:lamb-form}. This is different from all other sine-Gordon fields considered in this paper and will require a special treatment.

The string corresponding to the elementary solution is known as the Arutyunov-Frolov-Zamaklar magnon \cite{Arutyunov:2006gs}, i.e. the giant magnon with finite angular momentum. In \cite{Okamura:2006zv}, this string was named single spin helical string of type $(i)$. Type $(ii)$ corresponds to the doubled region.

\begin{figure}[t!]
\begin{center}
\begin{tabular}{p{75mm}p{68mm}}
\parbox{75mm}{\begin{flushright}\includegraphics[scale=0.53]{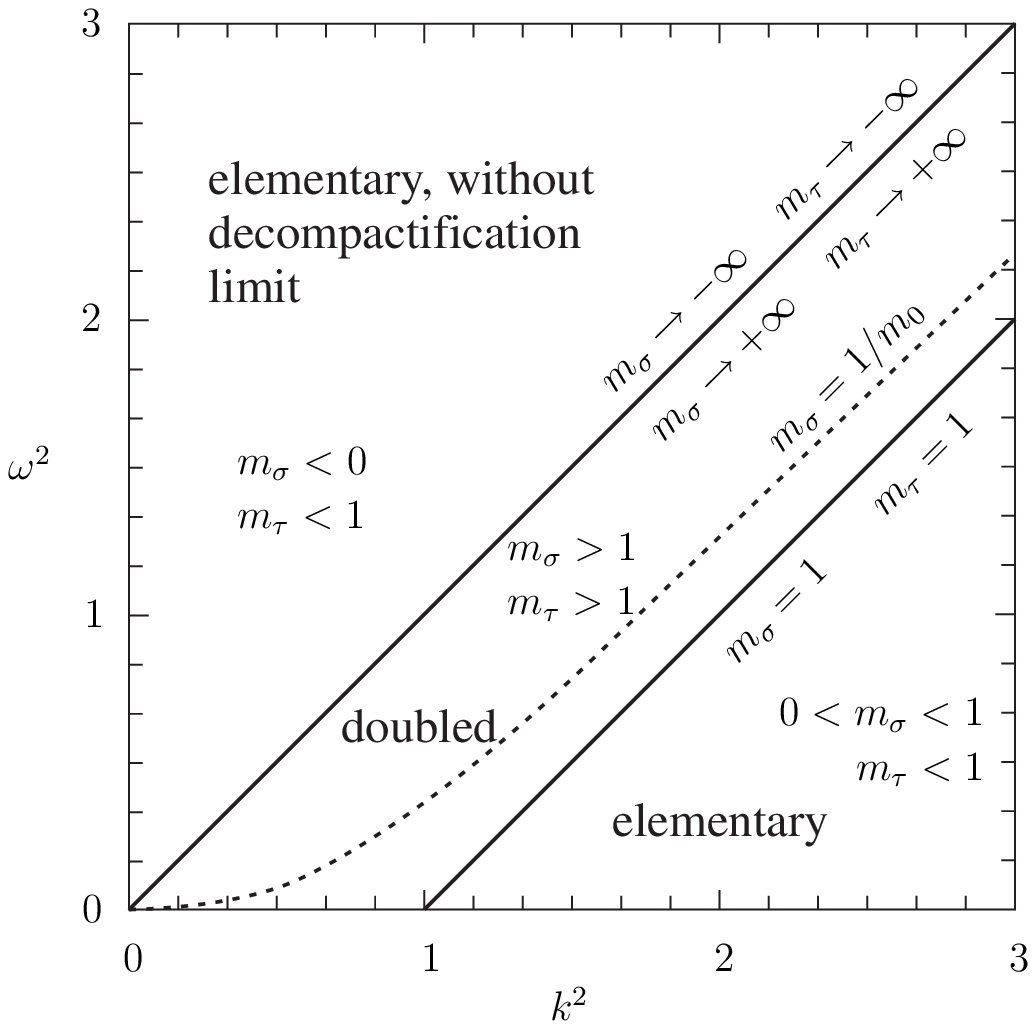}\end{flushright}} &
\parbox{68mm}{
(a) Fluxon oscillation \\[3mm]
Above the diagonal there is a region of elementary fuxon oscillations which is disconnected from the decompactification limit at $k^2 = \omega^2 + 1$.
} \\
\parbox{75mm}{\begin{flushright}\includegraphics[scale=0.53]{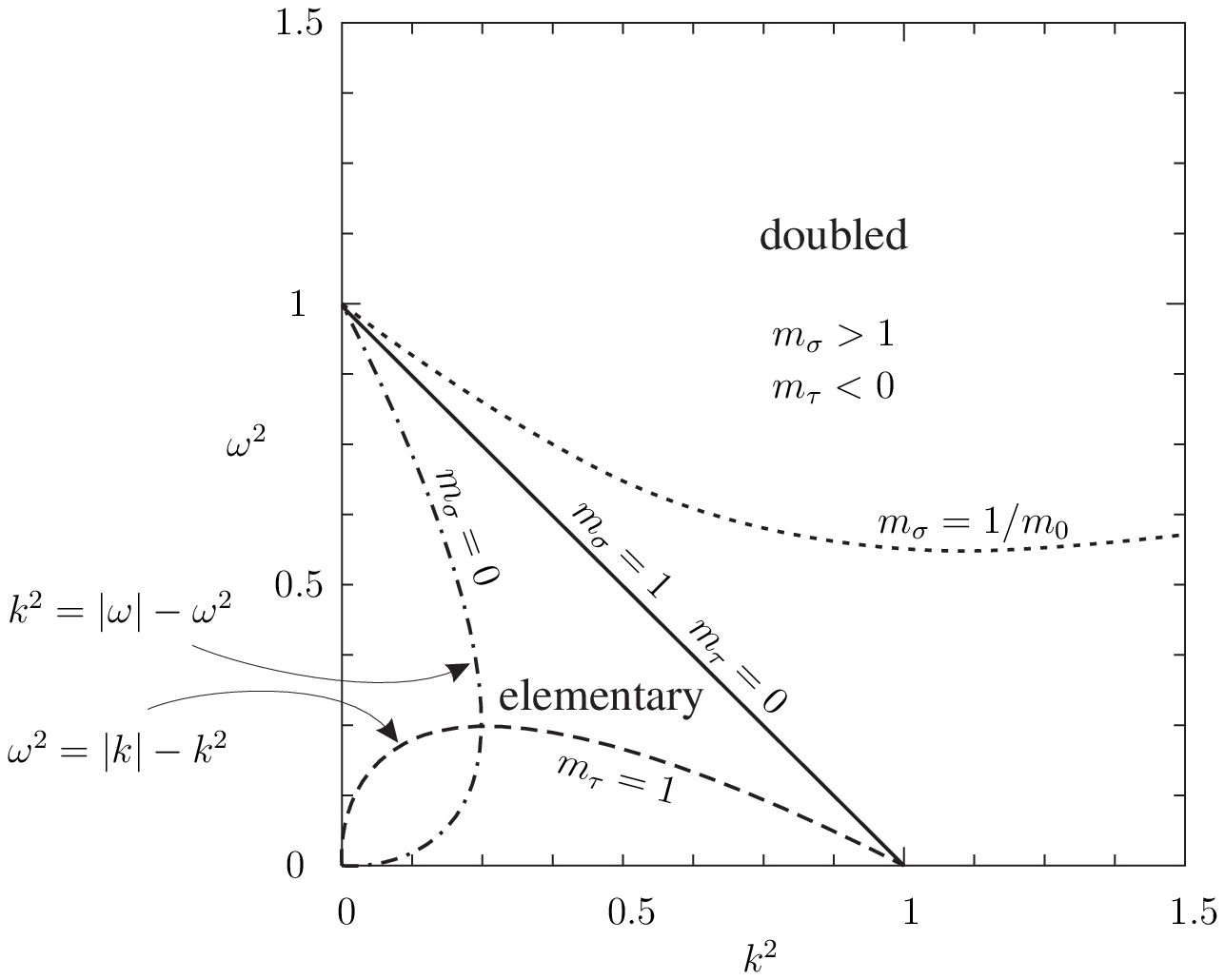}\end{flushright}} &
\parbox{68mm}{
(b) Fluxon breather \\[3mm]
The elementary region ($k^2+\omega^2\le1$) is subdivided into regions where $m_\sigma<0$ (left of dash-dotted line) and $0<m_\sigma<1$ (right of dash-dotted line) and into regions where $0<m_\tau<1$ (above dashed line) and $1<m_\tau$ (below dashed line).
} \\
\parbox{75mm}{\begin{flushright}\includegraphics[scale=0.53]{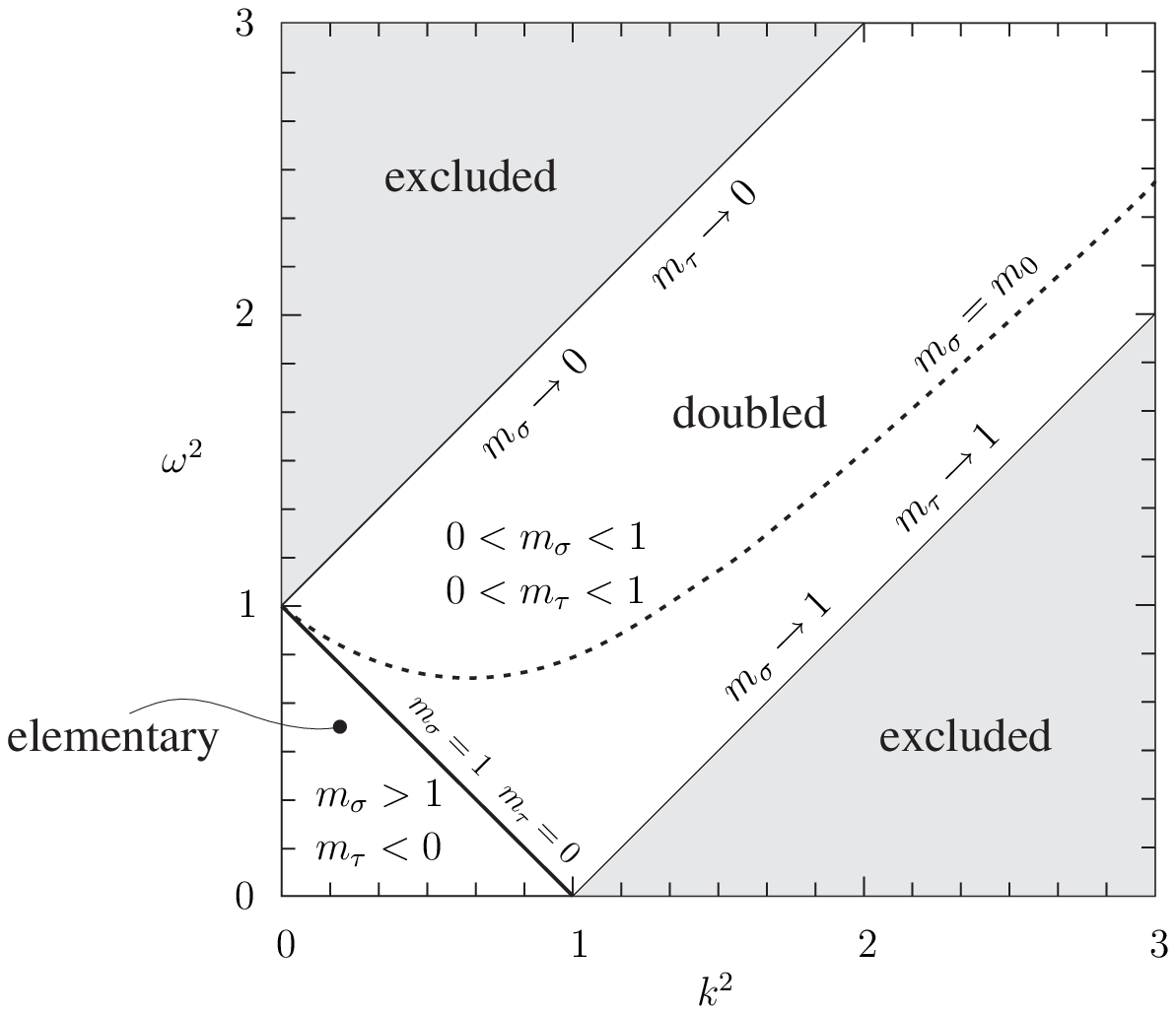}\end{flushright}} &
\parbox{68mm}{
(c) Plasmon breather \\[3mm]
The shaded regions and their boundaries are excluded because the solution would be imaginary. The plasmon breather is always periodic in time.
}
\end{tabular}
\caption{\textbf{Parameter space for periodic solutions.} The spatial and temporal periods become infinite long the lines where $m_\sigma=1$ and $m_\tau=1$, respectively. It will turn out that along the dotted lines, and in case (b) also along the dash-dotted line, the angular momentum of the associated string vanishes. ($m_0 \approx 0.826$, $1/m_0\approx 1.21$)}
\label{fig:parameter-spaces}
\end{center}
\end{figure}

\subsection{Fluxon oscillation}

The periodic generalization of the scattering solution \eqref{eqn:SG-scatt-decomp} is given by \cite{Costabile:1978}.
\be \label{eqn:SG-scatt-finite}
  \phi(\sigma,\tau) = 2 \arctan\bigsbrk{A \JacobiDN(k\sigma|m_\sigma)\JacobiSC(\omega \tau|m_\tau)} \; ,
\ee
where $k$ and $\omega$ are free parameters of the solution. The elliptic moduli and amplitude are determined by
\be
  m_\sigma = 1 - \frac{\omega^2}{k^2} \frac{1-k^2+\omega^2}{\omega^2-k^2}
  \comma
  m_\tau = 1 - \frac{k^2}{\omega^2} \frac{1-k^2+\omega^2}{\omega^2-k^2}
  \comma
  A = \frac{k}{\omega} \; .
\ee
Although \eqref{eqn:SG-scatt-finite} is a real solution for arbitrary real values of the parameters, we will restrict ourselves to $k,\omega>0$. In this way we avoid awkward case differentiations, and, if desired, results outside this region can be obtained by a reflection in $\sigma$ and/or $\tau$.

As in the case of a single kink train, there is an elementary region determined by $m_\sigma < 1$ and a doubled region determined by $m_\sigma > 1$. In fact, these conditions divide the parameter space $(k,\omega)$ in the exactly same way as before. A graphical representation of the parameter space is given in \figref{fig:parameter-spaces}(a). In the elementary region the solution is quasi-periodic and there is one kink and one anti-kink scattering off each other within one period, see \figref{fig:sol-scatt-finite-e}. In the doubled region the solution is strictly periodic and one period contains besides the two scattering kink anti-kink pair also their mirror image, see \figref{fig:sol-scatt-finite-d}.

The (quasi-)periods are given by
\be \label{eqn:L-scatt}
  L = \begin{cases}
      \frac{2}{k}  \, \EllipticK(m_\sigma) & \mbox{for $k^2\le\omega^2$ or $k^2\ge\omega^2+1$} \; , \\
      \frac{4}{k\sqrt{m_\sigma}}  \, \EllipticK\bigbrk{\tfrac{1}{m_\sigma}} & \mbox{for $\omega^2<k^2<\omega^2+1$} \; ,
      \end{cases}
\ee
and
\be \label{eqn:T-scatt}
  T = \begin{cases}
      \frac{2}{\omega}  \, \EllipticK(m_\tau) & \mbox{for $k^2\le\omega^2$ or $k^2\ge\omega^2+1$} \; , \\
      \frac{4}{\omega\sqrt{m_\tau}}  \, \EllipticK\bigbrk{\tfrac{1}{m_\tau}} & \mbox{for $\omega^2<k^2<\omega^2+1$} \; ,
      \end{cases}
\ee
which have the same functional form as in the kink train case. In the decompactification limit, $L,T\to\infty$, the parameters satisfy $k^2 = 1 + \omega^2$ and the solution reduces to the \eqref{eqn:SG-scatt-decomp} with the identification
\be \label{eqn:identify-scattering-decomp}
  k = \gamma
  \comma
  \omega = \gamma\beta
  \; .
\ee

By a shift
\be \label{eqn:shift-asol-to-sol-finte}
  k\sigma \to k\sigma + i\EllipticK'(m_\sigma)
  \comma
  \omega\tau \to \omega\tau + i\EllipticK'(m_\tau) \; ,
\ee
which is analogous to \eqref{eqn:shift-asol-to-sol}, one obtains the scattering of two kink trains given by
\be \label{eqn:SG-ss-scatt-finite}
  \phi(\sigma,\tau) = 2 \arctan\bigsbrk{A \JacobiCS(k\sigma|m_\sigma)\JacobiND(\omega \tau|m_\tau)} \; .
\ee

The strings constructed from the solutions \eqref{eqn:SG-scatt-finite} and \eqref{eqn:SG-ss-scatt-finite} describe the scattering of two giant magnons at finite angular momentum.

\subsection{Fluxon breather}

Like in the decompactified case, a breather solution can be obtained from the fluxon oscillation \eqref{eqn:SG-scatt-finite} by analytically continuing the frequency parameter $\omega\to i\omega$. Using the identity $\JacobiSC(iu|m)= i \JacobiSN(u|1-m)$, one obtains the fluxon breather \cite{Costabile:1978}
\be \label{eqn:SG-bound-finite}
  \phi(\sigma,\tau) = 2 \arctan\bigsbrk{A \JacobiDN(k\sigma|m_\sigma)\JacobiSN(\omega\tau|m_\tau)}
\ee
where the elliptic moduli and amplitude are now given by
\be \label{eqn:parameters-fluxon}
  m_\sigma = 1 - \frac{\omega^2}{k^2} \frac{1-k^2-\omega^2}{\omega^2+k^2}
  \comma
  m_\tau = \frac{k^2}{\omega^2} \frac{1-k^2-\omega^2}{\omega^2+k^2}
  \comma
  A = \frac{k}{\omega} \; .
\ee
As before we choose $k,\omega>0$ for simplicity and deduce results in other regimes by reflecting the coordinates $\sigma$ and/or $\tau$. The parameter space for this solution is very rich, cf.~\figref{fig:parameter-spaces}(b), and allows for many interesting special string solutions. In the elementary region the solution contains a bound soliton anti-soliton pair, \figref{fig:sol-bound-finite-e}, which received a mirror pair once one goes to the doubled region, \figref{fig:sol-bound-finite-d}. 

The solution is everywhere strictly periodic with periods
\be
  L = \begin{cases}
      \frac{2}{k}  \, \EllipticK(m_\sigma) & \mbox{for $k^2+\omega^2\le1$} \; , \\
      \frac{4}{k\sqrt{m_\sigma}}  \, \EllipticK\bigbrk{\tfrac{1}{m_\sigma}} & \mbox{for $k^2+\omega^2>1$} \; ,
      \end{cases}
\ee
and $T = \frac{4}{\omega} \Re(\EllipticK(m_\tau))$. Resolving the real part in this formula yields
\be
  T = \begin{cases}
      \frac{4}{\omega}  \, \EllipticK(m_\tau) & \mbox{for $\omega^2\ge k(1-k)$} \; , \\
      \frac{4}{\omega\sqrt{m_\tau}}  \, \EllipticK\bigbrk{\tfrac{1}{m_\tau}} & \mbox{for $\omega^2<k(1-k)$} \; .
      \end{cases}
\ee
Along the line $k^2 + \omega^2 = 1$ in parameter space the solution becomes decompactified, $L=\infty$, but remains $\tau$-periodic with period $T=\frac{2\pi}{\omega}$. The solution goes over into the breather solution \eqref{eqn:SG-bound-decomp} with the identification
\be \label{eqn:identify-breather-decomp}
  k = \gamma_a
  \comma
  \omega = \gamma_a a
  \; .
\ee

\subsection{Plasmon breather}

The fluxon breather is not the only periodic generalization of the breather on the infinite line. Another solution, with the same decompactification limit, is given by the plasmon breather \cite{Costabile:1978}
\be \label{eqn:SG-plasmon}
  \phi(\sigma,\tau) = 2 \arctan\bigsbrk{A \JacobiCN(k\sigma|m_\sigma)\JacobiCN(\omega\tau|m_\tau)} \; .
\ee
The elliptic moduli and the amplitude are related by
\be
  m_\sigma = \frac{(1+k^2)^2 - \omega^4}{4k^2}
  \comma
  m_\tau = \frac{k^4-(1-\omega^2)^2}{4\omega^2}
  \comma
  A = \pm \sqrt{\frac{1+k^2-\omega^2}{1-k^2+\omega^2}}
\ee
to the free parameters $k$ and $\omega$. Those must satisfy $\abs{k^2-\omega^2}<1$ for $\phi$ to be real. This results in a smaller parameter space than for the previous solutions, cf.~\figref{fig:parameter-spaces}(c). The plasmon breather is qualitatively very similar to the fluxon breather and we refer to the latter in \figref{fig:solitons} to get a visual impression.

The periods are
\be
  L = \begin{cases}
      \frac{2}{k\sqrt{m_\sigma}}  \, \EllipticK\bigbrk{\tfrac{1}{m_\sigma}} & \mbox{for $k^2+\omega^2<1$} \; , \\
      \frac{4}{k}  \, \EllipticK(m_\sigma) & \mbox{for $k^2+\omega^2\ge1$} \; ,
      \end{cases}
\ee
and
\be
  T = \frac{4}{\omega}  \, \EllipticK(m_\tau) \; .
\ee
As for the fluxon breather, the spatial period diverges for $k^2 + \omega^2 = 1$ while the temporal period stays finite and equal to $T=\frac{2\pi}{\omega}$. The solution goes also into the breather solution \eqref{eqn:SG-bound-decomp} with the same identification \eqref{eqn:identify-breather-decomp} of the parameters as in the fluxon breather case, but with an additional shift in $\tau$ by $\frac{\pi}{2\omega}$.

Given that the fluxon oscillation and the fluxon breather are related by analytic continuation in $\omega$, it is natural to ask what happens when we set $\omega \to i\omega$ in the plasmon breather \eqref{eqn:SG-plasmon}. Interestingly, one does not obtain a scattering solution, but instead recovers the plasmon breather itself with the $\tau$ coordinate shifted by $\frac{1}{\omega} \lrbrk{ \EllipticK(m_\tau) + i\EllipticK'(m_\tau) }$.

\section{From sine-Gordon to strings on $\Reals\times\Sphere^2$}
\label{sec:reconstruction}

In this section we reconstruct the target-space strings corresponding to the solutions of the sine-Gordon equation discussed in the previous section. The non-periodic sine-Gordon fields and the kink train lead to well known giant magnon solutions and are briefly treated here for completeness and as instructive examples. The periodic two-phase solutions lead to novel classical closed string solutions on $\Reals\times\Sphere^2$ describing two interacting giant magnons with finite angular momenta.

Technically we are facing the problem of inverting \eqref{eqn:defSGfield}. The derivatives in this mapping cause the string to depend non-locally on the sine-Gordon field. Due to this complication no general inverse map has been found so far though for previous related work in this context see \cite{Mikhailov:2005qv,Hofman:2006xt}.

Our approach is to utilize the formalism used in the theory of surfaces. In fact, the string target-space vector $\vec{n}(\sigma,\tau)$ parametrizes a patch on the unit sphere which in general overlaps with itself. In the chosen gauge, the coordinate $\tau$ also represents time. The coordinate lines which correspond to light-cone coordinates on the world sheet are what is known as a Chebyshev net \cite{Chebyshev:1946} in the mathematical literature. These nets are characterized by the condition that in any net quadrangle the opposite sides are equal; here this condition is contained in \eqref{eqn:constraints_lightcone}. The angle $2\phi$ between these coordinate lines determines the curvature of the surface. A surface on the unit sphere has constant Gaussian curvature $+1$ and the net angles satisfy the sine-Gordon \eqref{eqn:SGeq} as a consistency condition.

The general formalism works with the trihedron of the surface given by the orthonormal vectors $\{\vec{n},\vec{e}_1,\vec{e}_2\}$. On the sphere the coordinate vector $\vec{n}$ also serves as the unit normal vector. The fundamental equations for the trihedron are ($a,b=1,2$)
\begin{align}
  d\vec{e}_a & = \omega_a^b \, \vec{e}_b + \omega_a^3 \, \vec{n} \; , \label{eqn:fund-surface-1} \\
  d\vec{n}   & = \omega_3^b \, \vec{e}_b \; ,                         \label{eqn:fund-surface-2}
\end{align}
where we make the following choice for the connection on the sphere
\be \label{eqn:fund-coeffs}
  \omega_1^2 = \phi_\tau \, d\sigma + \phi_\sigma \, d\tau
  \comma
  \omega_1^3 = - \cos\phi \, d\tau
  \comma
  \omega_2^3 = - \sin\phi \, d\sigma
\ee
with $\omega_i^j = - \omega_j^i$ for $i,j=1,2,3$. From this one reads off the first partial derivatives of the basis vectors
\be \label{eqn:fund-eqn-1}
\matr{c}{\vec{e}_{1,\sigma} \\ \vec{e}_{2,\sigma} \\ \vec{n}_\sigma }
= \matr{ccc}{0 & \phi_\tau & 0 \\ -\phi_\tau & 0 & -\sin\phi \\ 0 & \sin\phi & 0}
\matr{c}{\vec{e}_1 \\ \vec{e}_2 \\ \vec{n}}
\ee
and
\be \label{eqn:fund-eqn-2}
\matr{c}{\vec{e}_{1,\tau} \\ \vec{e}_{2,\tau} \\ \vec{n}_\tau }
= \matr{ccc}{0 & \phi_\sigma & -\cos\phi \\ -\phi_\sigma & 0 & 0 \\ \cos\phi & 0 & 0}
\matr{c}{\vec{e}_1 \\ \vec{e}_2 \\ \vec{n}} \; .
\ee
The compatibility of the second derivatives follows from the sine-Gordon equation for $\phi$.

\paragraph{Reconstruction at fixed $\tau$.} We would like to keep the reconstruction as general as possible. That being said, in order to make progress we have to assume that the sine-Gordon field $\phi(\sigma,\tau)$ satisfies
\be \label{eqn:constancy-condition}
  \partial_\sigma \lrbrk{\frac{\phi_\tau}{\sin\phi}} \equiv 0 \; .
\ee
In fact this condition is obeyed by any solution of Lamb type \eqref{eqn:lamb-form}, in particular by all solutions discussed in the previous section except the kink train\footnote{This shows indirectly that the kink train cannot be cast into the factorized Lamb form. This in turn shows that \eqref{eqn:constancy-condition} cannot be a consequence of the sine-Gordon equation.}. For solutions subject to \eqref{eqn:constancy-condition}, the first half \eqref{eqn:fund-eqn-1} of the fundamental equations can be diagonalized explicitly because in this case all elements of the mixing matrix have the same $\sigma$ dependence. We find
\begin{align}
  \vec{e}_1(\sigma,\tau) & = - h(\tau) \vec{a}(\tau) + d(\tau) \bigsbrk{ \cos\alpha \, \vec{b}(\tau) + \sin\alpha \, \vec{c}(\tau) } 
                             \; , \label{eqn:e1x-phi} \\
  \vec{e}_2(\sigma,\tau) & = - \sin\alpha \, \vec{b}(\tau) + \cos\alpha \, \vec{c}(\tau)
                             \; , \label{eqn:e2x-phi} \\
  \vec{n}(\sigma,\tau)   & = d(\tau) \vec{a}(\tau) + h(\tau) \bigsbrk{ \cos\alpha \, \vec{b}(\tau) + \sin\alpha \, \vec{c}(\tau) }
                             \; , \label{eqn:nx-phi} 
\end{align}
where we have defined
\be
  h(\tau) := \frac{1}{\sqrt{ 1 + (\phi_\tau/\sin\phi)^2 }}
  \comma
  d(\tau) := h(\tau) \frac{\phi_\tau}{\sin\phi}
\ee
and
\be \label{eqn:def-alpha}
  \alpha(\sigma,\tau) := \frac{1}{h(\tau)} \int^\sigma\! \sin\phi(\sigma',\tau) \, d\sigma' \; .
\ee
The vectors $\vec{a}(\tau)$, $\vec{b}(\tau)$, $\vec{c}(\tau)$ are integration constants. They are required to be orthonormalized and independent of $\sigma$ but may depend on $\tau$. Their $\tau$-dependence can be determined from the second half \eqref{eqn:fund-eqn-2} of the fundamental equations. We do not need to introduce a further integration constant $\alpha_0(\tau)$ in \eqref{eqn:def-alpha} as this can be absorbed by a redefinition of $\vec{b}(\tau)$ and $\vec{c}(\tau)$. 

\begin{figure}[t]
\begin{center}
  \includegraphics[scale=0.6]{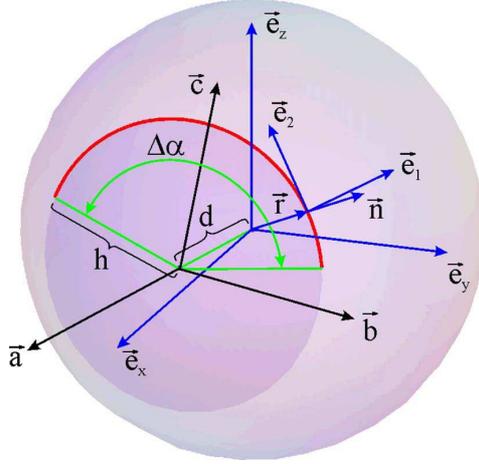}
\caption{\textbf{String reconstruction from sine-Gordon.} he global coordinate system is defined by the constant basis vectors $\{\vec{e}_x,\vec{e}_y,\vec{e}_z\}$. The position of the string in space-time is expressed in the time-dependent basis $\{\vec{a},\vec{b},\vec{c}\}$. The trihedron $\{\vec{e}_1,\vec{e}_2,\vec{n}\}$ is a set of basis vectors along the string.}
\label{fig:reconstruction}
\end{center}
\end{figure}

There is a very nice geometrical picture for \eqref{eqn:e1x-phi} to \eqref{eqn:nx-phi}. Consider the string $\vec{n}(\sigma,\tau)$ at some fixed time $\tau$. Eq.~\eqref{eqn:nx-phi} shows that the string stretches along a circular arc with central angle
\be
  \Delta\alpha(\tau) = \max_{\sigma} \alpha(\sigma,\tau) - \min_{\sigma} \alpha(\sigma,\tau) \; .
\ee
The arc lies in the $\{\vec{b}(\tau),\vec{c}(\tau)\}$ plane, has radius $h(\tau)$ and is centered at $d(\tau)\vec{a}(\tau)$. This is illustrated in \figref{fig:reconstruction}.

\paragraph{Complete reconstruction.} We are left with finding the $\tau$ dependence of the basis vectors $\vec{a}$, $\vec{b}$ and $\vec{c}$ from the second half of the fundamental equations. To this end we differentiate \eqref{eqn:e1x-phi}-\eqref{eqn:nx-phi} with respect to $\tau$ and set the result equal to \eqref{eqn:fund-eqn-2}. This yields
\be \label{eqn:2nd-fund-eqn}
  \matr{c}{\vec{a}_\tau \\ \vec{b}_\tau \\ \vec{c}_\tau} =
  \matr{ccc}{ 0 & M_{ab} & M_{ac} \\ -M_{ab} & 0 & M_{bc} \\ -M_{ac} & -M_{bc} & 0 }
  \matr{c}{\vec{a} \\ \vec{b} \\ \vec{c}}
\ee
with
\begin{align}
  M_{ab} & = \frac{h^2 \phi_{\sigma\sigma}}{\sin\phi} \cos\alpha + h \, \phi_\sigma \sin\alpha \; , \\
  M_{ac} & = \frac{h^2 \phi_{\sigma\sigma}}{\sin\phi} \sin\alpha - h \, \phi_\sigma \cos\alpha \; , \\
  M_{bc} & = \frac{h \, \phi_\tau\phi_\sigma}{\sin\phi} - \alpha_\tau \; .
\end{align}
Though it is not apparent, these matrix elements are independent of $\sigma$. It cannot be otherwise since the vectors $\vec{a}$, $\vec{b}$ and $\vec{c}$ do not depend on $\sigma$. The $\tau$ dependence, however, might be arbitrarily complicated. We proceed with the diagonalization of \eqref{eqn:2nd-fund-eqn} by defining the angle
\be \label{eqn:def-vartheta}
  \vartheta(\tau) := \arctan \frac{M_{ac}}{M_{ab}}
\ee
and introducing a rotated coordinate system through
\be
  \matr{c}{\vec{b}' \\ \vec{c}'} = \matr{cc}{\cos\vartheta & \sin\vartheta \\ -\sin\vartheta & \cos\vartheta } \matr{c}{\vec{b} \\ \vec{c}} \; .
\ee
In this coordinate system we have
\be \label{eqn:2nd-fund-eqn-rotated}
  \matr{c}{\vec{a}_\tau \\ \vec{b}'_\tau \\ \vec{c}'_\tau} = 
  \matr{ccc}{ 0 & \varphi_\tau & 0 \\ -\varphi_\tau & 0 & 0 \\ 0 & 0 & 0 }
  \matr{c}{\vec{a} \\ \vec{b}' \\ \vec{c}'}
\ee
with
\be \label{eqn:def-varphi}
  \varphi_\tau(\tau) := \frac{M_{ab}}{\cos\vartheta} = \pm \sqrt{ \biggbrk{\frac{h^2 \phi_{\sigma\sigma}}{\sin\phi}}^2 + \bigbrk{h\,\phi_\sigma}^2 } 
  \qquad\mbox{and}\qquad
  \varphi(\tau) = \int \varphi_\tau\, d\tau \; .
\ee
In order to show \eqref{eqn:2nd-fund-eqn-rotated} one has to make use of the sine-Gordon equation as well as the property \eqref{eqn:constancy-condition}. The solution of \eqref{eqn:2nd-fund-eqn-rotated}, rotated back to the original coordinate system is
\begin{align}
  \vec{a}(\tau) & = \sin\varphi \, \vec{a}_0 - \cos\varphi \, \vec{b}_0 \; , \nn \\
  \vec{b}(\tau) & = \cos\vartheta \bigsbrk{ \cos\varphi \, \vec{a}_0 + \sin\varphi \, \vec{b}_0  } - \sin\vartheta \, \vec{c}_0 \; , \label{eqn:sol-atbtct} \\
  \vec{c}(\tau) & = \sin\vartheta \bigsbrk{ \cos\varphi \, \vec{a}_0 + \sin\varphi \, \vec{b}_0  } + \cos\vartheta \, \vec{c}_0 \; , \nn
\end{align}
where $\{\vec{a}_0,\vec{b}_0,\vec{c}_0\}$ is some constant right-handed orthonormal basis. If we make the canonical choice $\{\vec{e}_x,\vec{e}_y,\vec{e}_z\}$, then the vectors in \eqref{eqn:sol-atbtct} are the ordinary basis vectors on the sphere
\be
  \vec{a}(\tau) = - \vec{e}_\varphi\bigbrk{\varphi(\tau)}
  \comma
  \vec{b}(\tau) = \vec{e}_\vartheta\bigbrk{\vartheta(\tau),\varphi(\tau)}
  \comma
  \vec{c}(\tau) = \vec{e}_r\bigbrk{\vartheta(\tau),\varphi(\tau)}
  \; ,
\ee
where the angles $\vartheta(\tau)$ and $\varphi(\tau)$ are determined only by the sine-Gordon field through \eqref{eqn:def-vartheta} and \eqref{eqn:def-vartheta}, respectively. When we plug this into \eqref{eqn:nx-phi}, we find the simple result
\begin{align}
  \vec{e}_1(\sigma,\tau) & = h \, \vec{e}_\varphi(\varphi) + d \, \vec{e}_\vartheta(\vartheta-\alpha,\varphi) \; , \\
  \vec{e}_2(\sigma,\tau) & = \vec{e}_r(\vartheta-\alpha,\varphi) \; , \\
  \vec{n}(\sigma,\tau)   & = - d \, \vec{e}_\varphi\bigbrk{\varphi} + h \, \vec{e}_\vartheta\bigbrk{\vartheta-\alpha,\varphi} \label{eqn:n-reconstructed-short}
\end{align}
with the functions $d=d(\tau)$, $h=h(\tau)$, $\alpha=\alpha(\sigma,\tau)$, $\vartheta=\vartheta(\tau)$, $\varphi=\varphi(\tau)$, as defined above. This solves the reconstruction of the string for any sine-Gordon solution that satisfies \eqref{eqn:constancy-condition} which includes all solutions of Lamb form. For practical usage, it is worth spelling out the string target-space vector \eqref{eqn:n-reconstructed-short} explicit as
\begin{equation} \label{eqn:n-reconstructed}
\vec{n}(\sigma,\tau)
             = \matr{c}{ d \sin\varphi + h \cos\varphi \cos(\alpha-\vartheta) \\
                        -d \cos\varphi + h \sin\varphi \cos(\alpha-\vartheta) \\
                                         h \sin(\alpha-\vartheta) } \; .
\end{equation}

\bigskip
From the constraints \eqref{eqn:virasoro_tau_sigma} and equation of motion \eqref{eqn:eom_tau_sigma} for the vector $\vec{n}$ we can deduce the following very non-trivial identities for the sine-Gordon field $\phi$ and the derived quantities $h$, $d$, $\alpha$, $\vartheta$ and $\varphi$:
\begin{align}
  & d \, \varphi_\tau \sin(\alpha-\vartheta) = h \, (\alpha_\tau-\vartheta_\tau) \; , \label{eqn:identity-1} \\
  & h \, \varphi_\tau \cos(\alpha-\vartheta) - d_\tau = h \cos\phi \; , \label{eqn:identity-2}\\
  & d \, \varphi_\tau \cos(\alpha-\vartheta) + h_\tau = d \cos\phi \; . \label{eqn:identity-3}
\end{align}
The latter two identities are related by the property $h^2+d^2=1$. From the $\partial_\tau$-part of the fundamental equations \eqref{eqn:fund-eqn-2}, we can derive two further identities
\be \label{eqn:identity-4}
  d \, \phi_\sigma = \alpha_\tau-\vartheta_\tau
  \comma
  h \, \phi_\sigma = \varphi_\tau \sin(\alpha-\vartheta) \; ,
\ee
which are related by \eqref{eqn:identity-1}. All of the above identities are ultimately a consequence of the sine-Gordon equation and the assumed property \eqref{eqn:constancy-condition}, though they are hard to verify in a direct way.

\paragraph{Angular momentum.} The angular momentum of the string in target space is given by
\be \label{eqn:J-from-SG}
  \vec{\reJ} = \int\!d\sigma\: \vec{n}\times\vec{n}_\tau
             = \int\!d\sigma\: \cos\phi \, \vec{e}_2
             = \int\!d\sigma\: \cos\phi \, \bigsbrk{ - \sin\alpha \, \vec{b}(\tau) + \cos\alpha \, \vec{c}(\tau) }
             \; .
\ee
Since $\vec{\reJ}$ is conserved, we can compute it at some $\tau=\tau_0$ that is most convenient. Moreover, we can rotate our coordinate system such that at $\tau=\tau_0$ the vectors $\vec{b}(\tau_0)$ and $\vec{c}(\tau_0)$ point into a preferred direction, e.g. such that
\be \label{eqn:J-formula}
  \vec{\reJ} = \int\!d\sigma\: \cos\phi \, \matr{c}{ \sin\alpha \\ \cos\alpha \\ 0 } \; .
\ee
This formula assumes a different coordinate system for any time $\tau$, but that does not matter for the modulus $\abs{\vec{\reJ}}$. In this way, we avoid having to use the full expressions \eqref{eqn:sol-atbtct} and can compute the angular moment even without having done the full reconstruction.

\subsection{Single magnon at infinite $J$}

\begin{figure}[ht]
\begin{center}
\subfigure[String profile]{\includegraphics[width=44mm]{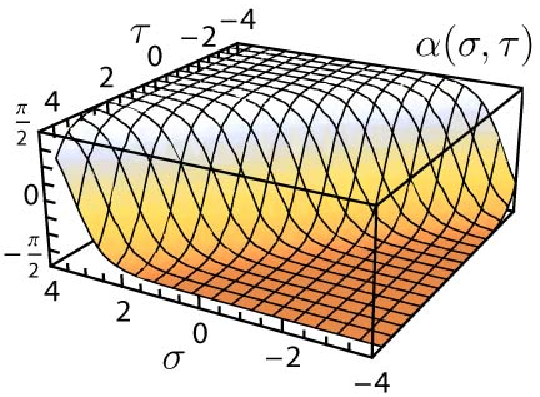}} \hspace{10mm}
\subfigure[Target-space string]{\hspace{5mm}\includegraphics[width=34mm]{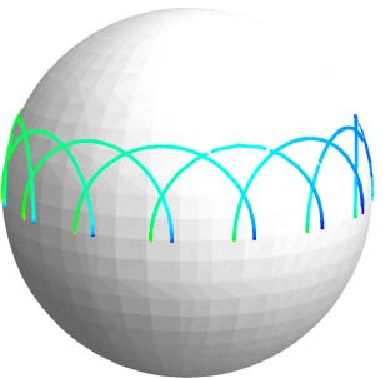}\hspace{5mm}}
\caption{\textbf{Single giant magnon.} These plots show the solution for $\beta=0.9$. Fig.~(a) shows the changing profile $\alpha(\sigma,\tau)$ of the string and Fig.~(b) depicts the string in target space at various fixed times of distance $\Delta\tau=0.4$. The coloring encodes the $\sigma$-coordinate along the string. From the way the color changes, one can see that one end of the string is stretched while the other is compressed. From within a reference frame that rotates together with the string around the sphere, this looks like a forbidden longitudinal motion of string bits. But actually this stretching and compressing is a consequence of the fact that there is \emph{no} longitudinal motion in the rest frame of the sphere.}
\label{fig:mag-single-decomp-analysis}
\end{center}
\end{figure}

We begin by applying the general reconstruction formulas derived above to the single kink solution \eqref{eqn:SG-single-decomp}. We readily find that the string radius is constant, $h(\tau) \equiv 1/\gamma$, and hence has constant distance $d(\tau) \equiv -\beta$ from the center of the sphere. The string profile is described by 
\be
  \alpha(\sigma,\tau) = 2\arctan\tanh\frac{\Gamma}{2}
  \qquad\mbox{and}\qquad
  \vartheta(\tau)     = -\frac{\pi}{2}
  \; ,
\ee
where we have introduced the notation $\Gamma(\sigma,\tau) = \gamma(\sigma-\beta \tau)$. The string's motion around the sphere is simply $\varphi(\tau) = \tau$. Plugging these functions into \eqref{eqn:n-reconstructed} we find the reconstructed string as
\be
\vec{n}(\sigma,\tau) = \matr{c}{
   - \beta \sin \tau - \tfrac{1}{\gamma} \, \cos \tau \, \tanh\Gamma \\
     \beta \cos \tau - \tfrac{1}{\gamma} \, \sin \tau \, \tanh\Gamma \\
                       \tfrac{1}{\gamma} \,              \sech\Gamma
} \; ,
\ee
which is, of course, the giant magnon of Hofman and Maldacena \cite{Hofman:2006xt} pictured in \figref{fig:mag-single-decomp-analysis}.

Using $\sin\alpha = \tanh\Gamma$, $\cos\alpha = \sech\Gamma$ and $\cos\phi = - \tanh\Gamma$ in the formula \eqref{eqn:J-formula}, we find for the components of the angular momentum
\be
  \vec{\reJ} = \int\!d\sigma\: \matr{c}{\tanh^2\Gamma \\ \tanh\Gamma\,\sech\Gamma \\ 0}
             = \matr{c}{ \eval{ \sigma }_{-\infty}^\infty - \tfrac{2}{\gamma} \\ 0 \\ 0} \; .
\ee
Hence, the modulus of the angular momentum can be written as
\be \label{eqn:J-mag-single-decomp}
  \reJ \equiv \abs{\vec{\reJ}} = L - \frac{2}{\gamma} \; ,
\ee
which is divergent due to the decompactified world-sheet $L = \infty$.

\subsection{Magnon--anti-magnon scattering at infinite $J$}

Reconstructing the string corresponding to the soliton scattering solution \eqref{eqn:SG-scatt-decomp} involves slightly more complicated expressions but is rather straightforward. Radius and position of the string obey
\be
  h(\tau) = \frac{1}{\sqrt{1+\beta^2\gamma^2\coth^2\gamma\beta \tau}}
  \comma
  d(\tau) = \frac{\beta\gamma\coth\gamma\beta \tau}{\sqrt{1+\beta^2\gamma^2\coth^2\gamma\beta \tau}}
  \; .
\ee
The string describes a full circle in target space where the string bits are distributed according to $\vartheta(\tau) = 0$ and
\be
  \alpha(\sigma,\tau) = 2\sign(\sinh\gamma\beta\tau) \arctan\frac{\beta\sinh\gamma \sigma}{\sqrt{\beta^2 + \sinh^2\gamma\beta \tau}} \; .
\ee
The motion around the sphere follows from the integration of $\varphi_\tau(\tau) = -\gamma^2 h^2(\tau)$ and reads
\be
  \varphi(\tau) = -\tau + \arctan \frac{\tanh\gamma\beta \tau}{\beta\gamma} + \pi \theta(\tau) \; .
\ee
Inserting these functions into the general position vector \eqref{eqn:n-reconstructed} gives the reconstructed target-space string, which represents a magnon with momentum $p_1$ scattering off an anti-magnon with momentum $p_2=2\pi-p_1$. This is a special case of the string solution for arbitrary momenta $p_1$ and $p_2$ found in \cite{Spradlin:2006wk} by the dressing method.

\begin{figure}[t]
\begin{center}
\subfigure[String radius]{\includegraphics[width=44mm]{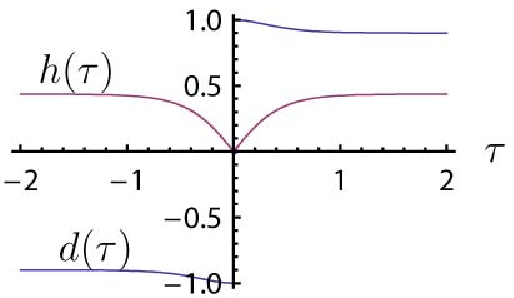}} \hspace{5mm}
\subfigure[Azimuthal angle]{\includegraphics[width=44mm]{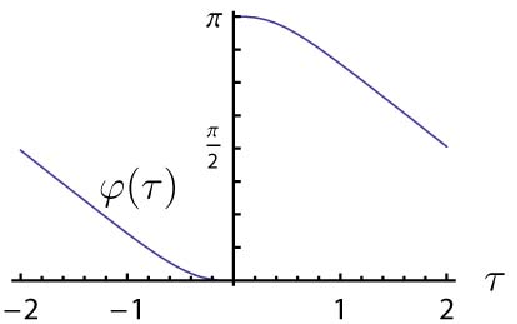}} \hspace{5mm}
\subfigure[String profile]{\includegraphics[width=44mm]{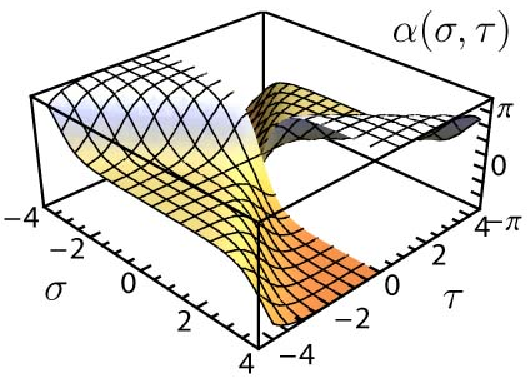}}
\caption{\textbf{Magnon--anti-magnon scattering at infinite $J$.} These plots visualize the solution for $\beta=0.9$.}
\label{fig:mag-amag-decomp-analysis}
\end{center}
\end{figure}
\begin{figure}[t]
\begin{center}
\subfigure[Magnon--anti-magnon scattering]{\hspace{12mm}\includegraphics[width=34mm]{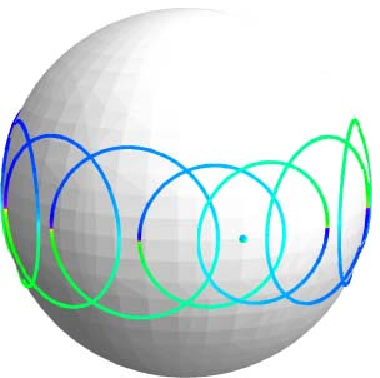}\hspace{8mm}
\label{fig:mag-amag-decomp-spacetime}}
\subfigure[Magnon--magnon scattering]{\hspace{8mm}\includegraphics[width=34mm]{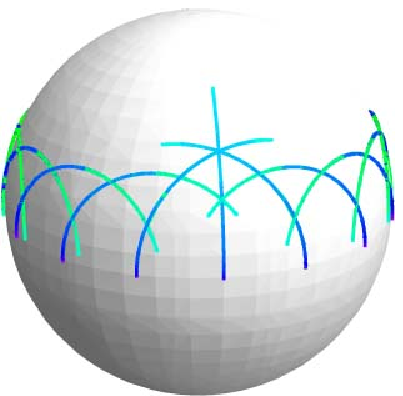}\hspace{8mm}
\label{fig:mag-mag-decomp-spacetime}}
\subfigure[Magnon breather]{\hspace{8mm}\includegraphics[width=34mm]{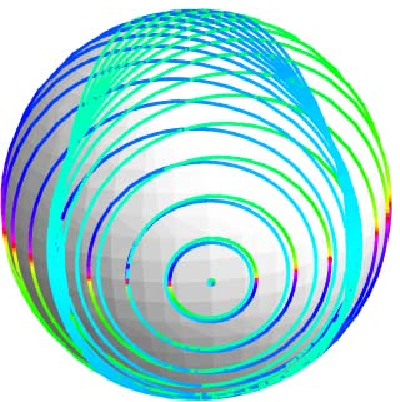}\hspace{4mm}
\label{fig:mag-bound-decomp-spacetime}}
\caption{\textbf{Two-magnon solutions at infinite $J$.} Shown are the strings in target space at various times of constant distance. The magnon--anti-magnon scattering solution is a circular string that spins around the equator and contracts to a point once at $\tau=0$. The magnon--magnon scattering solution is a folded string in the scape of a semi-circle. Once during its motion around the sphere, the forward ``endpoint'' detaches from the equator and it flips over to the back. The magnon breather solution is again a circular string that periodically shrinks to a point and in between sweeping over the entire sphere while progressing in azimuthal direction.}
\label{fig:two-mag-decomp}
\end{center}
\end{figure}

We have plotted the above functions in \figref{fig:mag-amag-decomp-analysis} and the target-space string in \figref{fig:mag-amag-decomp-spacetime} in order to describe some reoccurring features. The functions $d(\tau)$, $\varphi(\tau)$, and $\alpha(\sigma,\tau)$ are discontinuous across $\tau=0$, the time when the string radius vanishes, $h(0)=0$. Nevertheless, the mapping $\vec{n}(\sigma,\tau)$ from the world sheet to the target space is continuous everywhere. The discontinuities in $d$ and $\varphi$ exactly compensate each other, i.e. the sign flip in $d$ reflects the string along the axis defined by $\vec{a}$ (see \figref{fig:reconstruction}) while the jump from $0$ to $\pi$ in $\varphi$ rotates this axis by $180^\circ$. The discontinuity in the profile $\alpha$ does not harm either since it happens when the string has shrunken to a point. The result of all these discontinuities is an inversion of the string, which we have tried to indicate by coloring the string in \figref{fig:mag-amag-decomp-spacetime}. This inversion is necessary to preserve the angular momentum of the string.

By means of the formula \eqref{eqn:J-formula} for the angular momentum, we compute
\be \label{eqn:J-mag-amag-decomp}
  \reJ = L - \frac{4}{\gamma} \; ,
\ee
where again $L = \infty$. Being a two-magnon solution, the difference $\reJ-L$ is twice as large as for the single magnon \eqref{eqn:J-mag-single-decomp}.

\subsection{Magnon--magnon scattering at infinite $J$}

The string corresponding to the soliton--soliton scattering solution \eqref{eqn:SG-ss-scatt-decomp} is essentially the complex shift \eqref{eqn:shift-asol-to-sol} of the previous case. Therefore, we merely present the solution for radius and distance
\be
  h(\tau) = \frac{1}{\sqrt{1+\beta^2\gamma^2\tanh^2\gamma\beta\tau}}
  \comma
  d(\tau) = \frac{\beta\gamma\tanh\gamma\beta\tau}{\sqrt{1+\beta^2\gamma^2\tanh^2\gamma\beta\tau}} \; ,
\ee
string profile
\begin{align}
  \vartheta(\tau) = 0
  \comma
  \alpha(\sigma,\tau) = 2\arctan\frac{\beta\cosh\gamma\sigma}{\sqrt{\cosh^2\gamma\beta\tau-\beta^2}} \; ,
\end{align}
and the azimuthal motion
\be
  \varphi_\tau(\tau) = -\gamma^2 h^2(\tau)
  \quad\Rightarrow\quad
  \varphi(\tau) = -\tau - \arctan \gamma\beta \tanh\gamma\beta\tau \; .
\ee
The angular momentum is the same expression \eqref{eqn:J-mag-amag-decomp} as above. We note that these functions do not posses any discontinuities and refer to \figref{fig:mag-mag-decomp-spacetime} for a space-time picture.

\subsection{Magnon breather at infinite $J$}

Very similar to the previous cases is also the string corresponding to the soliton breather solution \eqref{eqn:SG-bound-decomp}. For completeness we note the component functions
\begin{align}
 &h(\tau) = \frac{1}{\sqrt{1+a^2\gamma_a^2\cot^2\gamma_a a \tau}}
  \comma
  d(\tau) = \frac{a\gamma_a\cot\gamma_a a \tau}{\sqrt{1+a^2\gamma_a^2\cot^2\gamma_a a \tau}} \; ,
  \\
 &\vartheta(\tau) = 0
  \comma
  \alpha(\sigma,\tau) = 2\sign(\sin \gamma_a a \tau) \arctan\frac{a\sinh\gamma_a \sigma}{\sqrt{a^2 + \sin^2\gamma_a a \tau}} \; ,
  \\
 &\varphi_\tau(\tau) = -\gamma_a^2 h^2(\tau)
  \quad\Rightarrow\quad
  \varphi(\tau) = -\tau + \arctan \frac{\tan\gamma_a a \tau}{a\gamma_a} + \pi \left\lfloor\frac{4\tau}{T}\right\rfloor
\end{align}
and the infinite angular momentum can be written as
\be
  \reJ = L - \frac{4}{\gamma_a} \; .
\ee
The string is plotted and described in \figref{fig:two-mag-decomp}.

\subsection{Magnon--anti-magnon scattering at finite $J$}

Now we turn to the reconstruction of novel closed string solutions with finite angular momenta based on the periodic two-phase sine-Gordon fields. The discussion of the periodic one-phase solution will be postponed until \secref{sec:reconstruction-mag-single} because it is not of Lamb form so that the general reconstruction formulas do not apply.
 
We begin with the fluxon oscillation \eqref{eqn:SG-scatt-finite}.  The reconstruction proceeds exactly as in the decompactified cases, different only in that it is technically more demanding because of the occurrence of elliptic functions. The functions relevant for the computation of the angular momentum are found to be
\begin{align}
  & h(\tau) = \frac{1}{\sqrt{1+\omega^2 \JacobiDS^2\omega\tau \JacobiNC^2\omega\tau}}
  \comma
  d(\tau) = \frac{\omega \JacobiDS\omega\tau \JacobiNC\omega\tau}{\sqrt{1+\omega^2 \JacobiDS^2\omega\tau \JacobiNC^2\omega\tau}}
  \\
  & \alpha(\sigma,\tau) = 2\sign(\JacobiSC\omega\tau) \arctan\frac{\sqrt{\omega^2+k^2(1-m_\sigma)\JacobiSC^2\omega\tau}\JacobiSC k \sigma}{\sqrt{\omega^2+k^2\JacobiSC^2\omega\tau}} \label{eqn:alpha-mag-amag-finite} \; ,
\end{align}
where we have omitted the elliptic moduli for notational brevity. If not stated otherwise, all elliptic functions with arguments $k\sigma$ and $\omega\tau$ are understood to have moduli $m_\sigma$ and $m_\tau$, respectively.

The branches of the $\arctan$ in \eqref{eqn:alpha-mag-amag-finite} have to be chosen appropriately. We place the branch cuts along the imaginary axis outside the unit circle. In the doubled region we can choose $-\pi/2 \le \arctan(\ldots) \le \pi/2$ for all $\sigma$, but in the elementary regions we have to define
\be
  (n-\half)\pi < \arctan(\ldots) \le (n+\half)\pi
  \quad\mbox{for}\quad
  (n-\half)L < \sigma \le (n+\half)L \; ,
\ee
such that $\alpha(\sigma,\tau)$ is a smooth function of $\sigma$.

Carrying out the complete reconstruction we find $\vartheta(\tau) = 0$ and
\be \label{eqn:varphi_t-fluxon-finite}
  \varphi_\tau(\tau) = - m_\sigma \, k^2 \, h^2(\tau) \; ,
\ee
which can be explicitly integrated. Using the main branches of the elliptic functions, $\varphi(\tau)$ can be expressed as 
\begin{equation}
\begin{split}
  \varphi(\tau) = \frac{\sign(\JacobiDN\omega\tau)}{\omega(k^2-\omega^2)} \biggsbrk{ &
  k^2 \, \EllipticPi\lrbrk{1-\tfrac{k^2}{\omega^2} ,
                           \am\omega\tau \big{|}
                           m_\tau } \\
  &
  - (k^2-\omega^2)(1-k^2+\omega^2) \, \EllipticPi\lrbrk{\tfrac{1}{k^2-\omega^2} ,
                                                        \am\omega\tau \big{|}
                                                        m_\tau} \\
  &
  - \omega\lrbrk{\omega^2+(k^2-\omega^2)^2} \, \tau \, }
  + \pi \left\lfloor \tfrac{\tau}{T/2} \right\rfloor \; ,
\end{split}
\end{equation}
in the elementary region, and as
\begin{equation}
\begin{split}
  \varphi(\tau) = &\ \frac{1}{\omega(k^2-\omega^2)\sqrt{m_\tau}} \biggsbrk{
  k^2 \, \EllipticPi\lrbrk{\lrbrk{1-\tfrac{k^2}{\omega^2}}\tfrac{1}{m_\tau} ,
                        \am(\sqrt{m_\tau}\,\omega\tau|\tfrac{1}{m_\tau}) \big{|}
                        \tfrac{1}{m_\tau} } \\
  & \qquad
  -(k^2-\omega^2)(1-k^2+\omega^2) \, \EllipticPi\lrbrk{\tfrac{1}{k^2-\omega^2}\tfrac{1}{m_\tau} ,
                                                    \am(\sqrt{m_\tau}\,\omega\tau|\tfrac{1}{m_\tau}) \big{|}
                                                    \tfrac{1}{m_\tau}} \\
  & \qquad
  - \omega\lrbrk{\omega^2+(k^2-\omega^2)^2}\sqrt{m_\tau} \: \tau \, }
  + \pi \left\lfloor \tfrac{\tau}{T/2} \right\rfloor
\end{split}
\end{equation}
in the doubled region.

\begin{figure}[ht]
\begin{center}
\subfigure[Magnon--anti-magnon scattering (elementary region with $k=2.29$ and $\omega=2.06$)]{
  \hspace{5mm}\includegraphics[width=34mm]{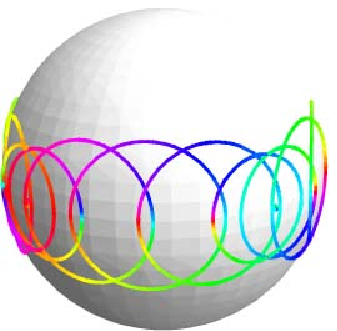}\hspace{5mm}
  \label{fig:mag-amag-finite-e-spacetime}} \hspace{6mm}
\subfigure[Magnon--magnon scattering (elementary region with $k = 2.29$ and $\omega = 2.06$)]{
  \hspace{5mm}\includegraphics[width=34mm]{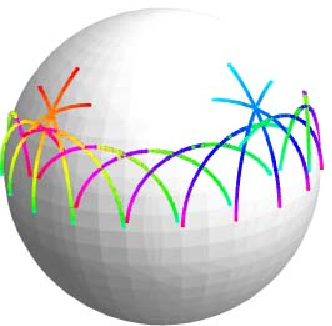}\hspace{5mm}
  \label{fig:mag-mag-finite-e-spacetime}} \hspace{6mm}
\subfigure[Fluxonic magnon breather (elementary region with $k = 0.83$ and $\omega = 0.55$)]{
  \hspace{5mm}\includegraphics[width=34mm]{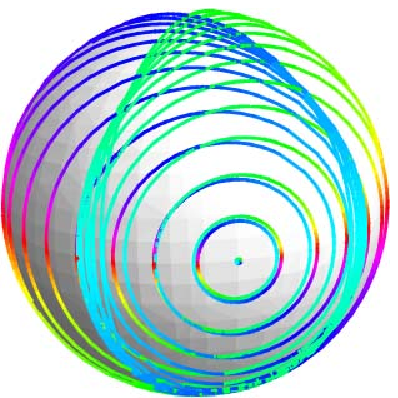}\hspace{5mm}
  \label{fig:flux-finite-e-spacetime}} \\
\subfigure[Magnon--anti-magnon scattering (doubled region with $k=2.405$ and $\omega=2.188$)]{
  \hspace{5mm}\includegraphics[width=34mm]{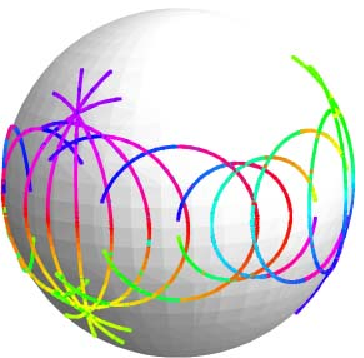}\hspace{5mm}
  \label{fig:mag-amag-finite-d-spacetime}} \hspace{6mm}
\subfigure[Magnon--magnon scattering (doubled region with $k = 2.405$ and $\omega = 2.188$)]{
  \hspace{5mm}\includegraphics[width=34mm]{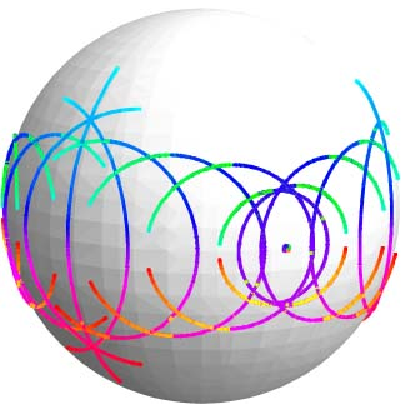}\hspace{5mm}
  \label{fig:mag-mag-finite-d-spacetime}} \hspace{6mm}
\subfigure[Fluxonic magnon breather (doubled region with $k = 0.90$ and $\omega = 0.89$)]{
  \hspace{5mm}\includegraphics[width=34mm]{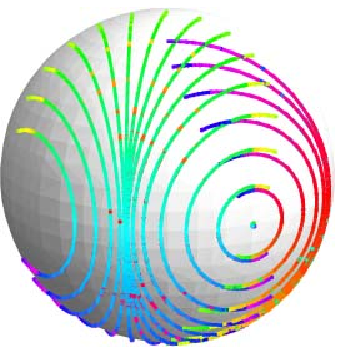}\hspace{5mm}
  \label{fig:flux-finite-d-spacetime}}
\caption{\textbf{Two-magnon solutions at finite $J$.} The first row shows representative solutions in the elementary region and the second row shows their counterparts in the doubled region. As before, the coloring indicates the dependence on the spatial coordinate $\sigma$ and appears non-smooth when the string is folded. The finite-$J$ scattering solutions (a) and (b) are essentially time periodic generalizations of the infinite-$J$ versions in {\protect\figref{fig:two-mag-decomp}}. The finite-$J$ breather (c) is very similar to the infinite-$J$ breather, but its moduli space much bigger, including e.g. the circular pulsating string. In the doubled regime, all finite-$J$ solutions have the same qualitative time evolution: a folded string that on its way around the equator periodically contracts to a point and stretches out maximally.}
\label{fig:mag-amag-finite-analysis}
\end{center}
\end{figure}

Figures \ref{fig:mag-amag-finite-e-spacetime} and \ref{fig:mag-amag-finite-d-spacetime} visualize this solution in the two regions, respectively. Note that the string $\vec{n}(\sigma,\tau)$ is not $T$-periodic in a strict sense because of the motion in the $\varphi$ direction. During one period the string advances by an azimuthal angle of $\Delta\varphi = \varphi(T)$ which is in general not a multiple of $2\pi$. Disregarding this motion around the sphere, the string is periodic in $\tau$.

In the elementary region, within one period $T$ the string contracts twice to a point and expands twice to maximal radius
\be
  h_{\mathrm{max}} = \frac{1}{\sqrt{1+\lrbrk{1+\sqrt{1-m_\tau}}^2 \omega^2}} \; .
\ee
In the decompactification limit this becomes
\be
  h_{\mathrm{max,decomp}} = \frac{1}{\sqrt{1+\omega^2}} = \frac{1}{\gamma} \; ,
\ee
where we used the identification \eqref{eqn:identify-scattering-decomp}. Twice this maximal radius is the separation of the two points where the string crosses the equatorial plane. This separation is related to the momenta of the individual magnons  which make up the scattering state \cite{Hofman:2006xt} and is based on the fact that the magnons in the decompactification limit cleanly separate. In the periodic, finite-volume case this is not true and maximal radius does not seem to be directly related to the individual magnons momenta, in fact as the magnons never separate it is not clear that such a concept is completely unambiguous. We will postpone further discussion of this point until \secref{sec:quantization}.

\bigskip
For the computation of the angular momentum, we notice that $\cos\phi$ and $\cos\alpha$ are even functions in $\sigma$, while $\sin\alpha$ is odd. This implies that $\reJ = \abs{\reJ_2}$ is the only non-vanishing component. Computing this component at $\tau=0$ we can use $\cos\phi(\sigma,0) = 1$ and $\cos\alpha(\sigma,0) = 1-2\JacobiSN^2 k\sigma$, and obtain
\be
  \reJ = \int_0^L\!d\sigma\: \JacobiCN^2(2k\sigma|m_\sigma) \; ,
\ee
which can be evaluated explicitly to
\be \label{eqn:J-mag-amag}
  \reJ = \begin{cases}
            \frac{2}{k \abs{m_\sigma}} \Bigsbrk{(2-m_\sigma)\EllipticK(m_\sigma) - 2\EllipticE(m_\sigma)}
          & \mbox{for $k^2\le\omega^2$ or $k^2\ge\omega^2+1$} \; , \\[3mm]
            \frac{4}{k \sqrt{m_\sigma}} \Bigabs{ \EllipticK\bigbrk{\frac{1}{m_\sigma}} - 2\EllipticE\bigbrk{\frac{1}{m_\sigma}} }
          & \mbox{for $\omega^2<k^2<\omega^2+1$} \; .
         \end{cases}
\ee
We note that in the first case, which corresponds to the elementary region, the angular momentum $\reJ$ is always strictly positive while in the second case, i.e.\ the doubled region, it vanishes along the curve in parameter space where $m_\sigma = 1/m_0 = 1.210485\ldots$, see the dotted line in \figref{fig:parameter-spaces}(a).

It is interesting to realize that the angular momentum can vanish although the string is monotonically orbiting the sphere, $\varphi_\tau>0$, as can be seen from \eqref{eqn:varphi_t-fluxon-finite}. The contributions of the individual string bits to the total angular momentum cancel between the central part and the ends of the folded string. When the ends reach over the north and south poles of the sphere this compensation can also happen when the center of mass of the string moves. In fact, if the target space was e.g. a cylinder, such a phenomenon could not occur.

\subsection{Magnon--magnon scattering at finite $J$}

Reconstructing the string from the soliton--soliton scattering solution \eqref{eqn:SG-ss-scatt-finite} can be done with equal time and effort from scratch or by a shift by a quarter of the imaginary period according to \eqref{eqn:shift-asol-to-sol-finte} and being very cautious about branch cuts. We find
\begin{align}
  & h(\tau) = \frac{1}{\sqrt{1+ m_\tau^2 \omega^2 \JacobiCN^2\omega\tau\JacobiSD^2\omega\tau}}
  \comma
  d(\tau) = \frac{ m_\tau \omega \JacobiCN\omega\tau\JacobiSD\omega\tau}{\sqrt{1+m_\tau^2\omega^2 \JacobiCN^2\omega\tau  \JacobiSD^2\omega\tau}} \\
  & \alpha(\sigma,\tau) = 2\sign(m_\sigma) \arctan\frac{\sqrt{\omega^2-k^2(1-m_\sigma)\JacobiND^2\omega \tau} \, \JacobiND k \sigma}{\sqrt{k^2\JacobiND^2\omega\tau-\omega^2}}
\end{align}
and, as before, $\vartheta(\tau) = 0$ and $\varphi_\tau(\tau) = - m_\sigma \, k^2 \, h^2(\tau)$. In the elementary regions, the integral of $\varphi_\tau$ simplifies to
\begin{equation}
\begin{split}
  \varphi(\tau) = \frac{1}{\omega(k^2-\omega^2)} \biggsbrk{ &
  - k^2 \, \EllipticPi\lrbrk{\tfrac{\omega^2 m_\tau}{\omega^2-k^2} ,
                             \am\omega\tau \big{|}
                             m_\tau } \\
  &
  + (k^2-\omega^2)(1-k^2+\omega^2) \, \EllipticPi\lrbrk{(k^2-\omega^2)m_\tau ,
                                                        \am\omega\tau \big{|}
                                                        m_\tau}
  } \; ,
\end{split}
\end{equation}
and in the doubled region, it can be written as
\begin{equation}
\begin{split}
  \varphi(\tau) = &\ \frac{1}{\omega(k^2-\omega^2)\sqrt{m_\tau}} \biggsbrk{
  - k^2 \, \EllipticPi\lrbrk{\tfrac{\omega^2}{\omega^2-k^2} ,
                             \am(\sqrt{m_\tau}\,\omega\tau|\tfrac{1}{m_\tau}) \big{|}
                             \tfrac{1}{m_\tau} } \\
  & \qquad
  +(k^2-\omega^2)(1-k^2+\omega^2) \, \EllipticPi\lrbrk{k^2-\omega^2 ,
                                                       \am(\sqrt{m_\tau}\,\omega\tau|\tfrac{1}{m_\tau}) \big{|}
                                                       \tfrac{1}{m_\tau}} } \\
  & + \pi \left\lfloor \tfrac{\tau+T/4}{T/2} \right\rfloor \; .
\end{split}
\end{equation}
Figures \ref{fig:mag-mag-finite-e-spacetime} and \ref{fig:mag-mag-finite-e-spacetime} show representative strings in the elementary and doubled region, respectively.

The angular momentum of this solution is the same as for the magnon--anti-magnon scattering given by \eqref{eqn:J-mag-amag}.

\subsection{Magnon breathers at finite $J$}

There are two breather solutions in sine-Gordon theory on the circle, the fluxon \eqref{eqn:SG-bound-finite} and the plasmon \eqref{eqn:SG-plasmon}. Accordingly, we call the strings related to these solutions the `fluxonic magnon breather' and the `plasmonic magnon breather'.

\subsubsection{Fluxonic magnon breather}
\label{sec:reconstruction-fluxon-breather}

From the fluxon breather solution \eqref{eqn:SG-bound-finite} we derive the functions
\begin{align}
  & h(\tau) = \frac{1}{\sqrt{1+\omega^2 \JacobiDS^2\omega\tau \JacobiCN^2\omega\tau}}
  \comma
  d(\tau) = \frac{\omega \JacobiDS\omega\tau \JacobiCN\omega\tau}{\sqrt{1+\omega^2 \JacobiDS^2\omega\tau \JacobiCN^2\omega\tau}} \\
  & \alpha(\sigma,\tau) = 2\sign(\JacobiSN\omega\tau) \arctan\frac{\sqrt{\omega^2+k^2(1-m_\sigma)\JacobiSN^2\omega\tau} \, \JacobiSC k \sigma}{\sqrt{\omega^2+k^2\JacobiSN^2\omega\tau}} \; .
\end{align}
The branches of the $\arctan$ are to be chosen as in the magnon--anti-magnon scattering case. Also, we have again $\vartheta(\tau) = 0$ and $\varphi_\tau(\tau) = - m_\sigma \, k^2 \, h^2(\tau)$, but since $h(\tau)$ is different, we find a different integral. For $m_\tau < 1$, which covers the entire doubled region as well as the part of the elementary region where $\omega^2 > k - k^2$, we have the formula
\begin{equation}
\begin{split}
  \varphi(\tau)  & = \tfrac{1}{\omega} \Bigsbrk{
                     \EllipticPi(-\tfrac{k^2}{\omega^2},\am\omega\tau|m_\tau)
                   - \EllipticPi(-m_\tau\tfrac{\omega^2}{k^2},\am\omega\tau|m_\tau)
                   }
                   + \pi \left\lfloor \tfrac{\tau}{T/2} \right\rfloor \; ,
\end{split}
\end{equation}
and in the remaining part $\omega^2 < k - k^2$ with $m_\tau > 1$ we have
\begin{equation}
\begin{split}
  \varphi(\tau)  & = \tfrac{1}{\omega} \Bigsbrk{
                     \EllipticPi\lrbrk{\tfrac{k^2+\omega^2}{k^2+\omega^2-1},
                                       \am(\sqrt{m_\tau}\,\omega\tau|\tfrac{1}{m_\tau}) \big{|}
                                       \tfrac{1}{m_\tau} } \\
                 & \qquad - \EllipticPi\lrbrk{-\tfrac{\omega^2}{k^2},
                                       \am(\sqrt{m_\tau}\,\omega\tau|\tfrac{1}{m_\tau}) \big{|}
                                       \tfrac{1}{m_\tau} }
                   }
                   + \pi \left\lfloor \tfrac{\tau}{T/2} \right\rfloor \; .
\end{split}
\end{equation}

The angular momentum turns out to be the same functions of $m_\sigma$ as in the scattering case
\be
  \reJ = \begin{cases}
            \frac{2}{k \abs{m_\sigma}} \Bigsbrk{(2-m_\sigma)\EllipticK(m_\sigma) - 2\EllipticE(m_\sigma)}
          & \mbox{for $k^2+\omega^2\le1$} \; , \\[3mm]
            \frac{4}{k \sqrt{m_\sigma}} \Bigabs{ \EllipticK\bigbrk{\frac{1}{m_\sigma}} - 2\EllipticE\bigbrk{\frac{1}{m_\sigma}} }
          & \mbox{for $k^2+\omega^2>1$} \; .
         \end{cases}
\ee
though, of course, the regions are different as is the dependence of $m_\sigma$ on the parameters $k$ and $\omega$, see \eqref{eqn:parameters-fluxon}. As before the angular momentum vanishes in the doubled region for parameters such that $m_\sigma = 1/m_0 = 1.210485\ldots$. This curve is plotted in the parameter space diagram \figref{fig:parameter-spaces}(b).
 
Besides this, the angular momentum also vanishes along the curve $k^2=\omega(1-\omega)$ through the elementary region of parameter space where $m_\sigma=0$. This is a very interesting family of solutions which we consider more explicitly. While $h(\tau)$ and $d(\tau)$ are unchanged, the other functions simplify to
\be
  \alpha(\sigma,\tau) = 2 \sign(\JacobiSN\omega\tau) \, k \sigma
  \comma
  \varphi(\tau) = 0 \; ,
\ee
and the range of $\sigma$ becomes $0$ to $L=\frac{\pi}{k}$. By introducing a rescaled coordinate $\sigma'=2k\sigma$, we can eliminate $k$ from all formulas and find the explicit solution
\be \label{eqn:pulsating-string}
  \vec{n}(\sigma,\tau) = \frac{\sign(\JacobiSN\omega\tau)}{\sqrt{1+\omega^2 \JacobiDS^2\omega\tau \JacobiCN^2\omega\tau}}
  \matr{c}{
    \cos\sigma' \\
    -\omega \JacobiDS\omega\tau \JacobiCN\omega\tau \\
    \sin\sigma'
  }
\ee
with $\sigma'=0\ldots2\pi$. This solution describes a circular string that pulsates between two antipodes from east to west and in between sweeping the entire sphere. The period of one oscillation depends on $\omega$ via
\be
  T = \frac{4}{\omega} \, \EllipticK(m_\tau)
  \quad\mbox{with}\quad
  m_\tau = \lrbrk{\frac{1}{\omega}-1}^2
  \; .
\ee
For $\omega = \half$ this period becomes infinite. In this case the string wraps a great circle at $\tau\to\pm\infty$ and contracts once to a point on one side of the sphere at $\tau=0$.

In \secref{sec:dispersion-fluxon-breather} we show that the semi-classical energy spectrum of the pulsating string solution \eqref{eqn:pulsating-string} coincides with the results of \cite{Minahan:2002rc}.

\subsubsection{Plasmonic magnon breather}
\label{sec:reconstruction-plasmon-breather}

The plasmon breather solution \eqref{eqn:SG-plasmon} gives rise to the following functions
\begin{align}
  & h(\tau) = \frac{1}{\sqrt{1+\omega^2 \JacobiDC^2\omega\tau \JacobiSN^2\omega\tau}}
  \comma
  d(\tau) = \frac{-\omega \JacobiDC\omega\tau \JacobiSN\omega\tau}{\sqrt{1+\omega^2 \JacobiDC^2\omega\tau \JacobiSN^2\omega\tau}} \\
  & \alpha(\sigma,\tau) = 2\sign(\JacobiCN\omega\tau) \sign(A) \arctan\frac{\sqrt{m_\sigma + A^2(m_\sigma-1)\JacobiCN^2\omega\tau} \, \JacobiSD k\sigma}{\sqrt{1+A^2 \JacobiCN^2\omega\tau}}
\end{align}
and once more we have $\vartheta=0$ and $\varphi_\tau = - k^2 h^2(\tau)$. Using appropriate branches of the elliptic integrals, we can write
\begin{equation}
\begin{split}
  \varphi(\tau) =\frac{\sqrt{f}}{\omega} \biggsbrk{
   & \mp \sqrt{1-(k^2+\omega^2)^2} \, \Bigbrk{\EllipticPi(f g, u | f^2) - \EllipticPi(\tfrac{f}{g}, u | f^2)} \\
   & + \sqrt{1-(k^2-\omega^2)^2} \, \Bigbrk{\EllipticPi(f \ell, u | f^2) - \EllipticPi(\tfrac{f}{\ell}, u | f^2)} \\
   & + 2ik^2 \EllipticF(u|f^2) + \pi \left\lfloor \tfrac{\tau+T/4}{T/2} \right\rfloor
  } \; .
\end{split}
\end{equation}
where the upper sign applies to the elementary and the lower sign to the doubled region. The argument is defined as $u(\tau) = i \arcsinh\bigsbrk{f^{-1/2}\tan\bigbrk{\half\am(\omega \tau|m_\tau)}}$ and the parameters are
\begin{align}
  f & = \tfrac{1}{2\omega^2}\bigbrk{1+\omega^4-k^4+\sqrt{[(\omega^2+k^2)^2-1][(\omega^2-k^2)^2-1]} } \; , \\
  g & = \omega^2 + k^2 + \sqrt{(\omega^2+k^2)^2-1} \; , \\
  \ell & = \omega^2 - k^2 + \sqrt{(\omega^2-k^2)^2-1} \; .
\end{align}
As mentioned previously, the plasmon breather is qualitatively extremely similar to the fluxon breather. This similarity carries over to the reconstructed string solution. Therefore, we refrain from drawing the plasmonic magnon breather separately and refer to figures \figref{fig:flux-finite-e-spacetime} and \figref{fig:flux-finite-d-spacetime} showing the fluxonic analog.

The computation of the angular momentum is simplest at $\tau=\quarter T = \frac{1}{\omega}\EllipticK(m_\tau)$ as we have
\be
  \cos\phi(\sigma,T/4) = 1
  \comma
  \alpha(\sigma,T/4) = \pm 2\arctan\sqrt{m_\sigma} \JacobiSD k\sigma \; .
\ee
Since $\alpha$ is an odd function of $\sigma$, the only contribution to the angular momentum is given by
\be
  \reJ = \lrabs{ \int_{0}^{L}\!d\sigma\: \cos\alpha }
       = \lrabs{ \int_{0}^{L}\!d\sigma\: (2\JacobiDN^2 k\sigma - 1) }
\ee
and integration yields
\be
  \reJ = \begin{cases}
            \frac{2}{k\sqrt{m_\sigma}} \Bigsbrk{ (2m_\sigma-1)\EllipticK\bigbrk{\frac{1}{m_\sigma}} -2 m_\sigma \EllipticE\bigbrk{\frac{1}{m_\sigma}} }
          & \mbox{for $k^2+\omega^2<1$} \; , \\[3mm]
            \frac{4}{k} \Bigabs{2\EllipticE(m_\sigma) - \EllipticK(m_\sigma)}
          & \mbox{for $k^2+\omega^2\ge1$} \; .
         \end{cases}
\ee
The angular moment is strictly positive in the entire elementary region and vanishes in the doubled region along the curve in parameter space where $m_\sigma = m_0 = 0.826114\ldots$, see the dotted line in \figref{fig:parameter-spaces}(c).

\subsection{Single magnon at finite $J$}
\label{sec:reconstruction-mag-single}

For completeness, and as it necessary to interpret the two magnon solutions, let us give a brief summary of the finite-$J$ magnon in conformal gauge as originally described by \cite{Arutyunov:2006gs} but using our notations.

As the underlying sine-Gordon kink train \eqref{eqn:SG-single-finite} is not of the Lamb form \eqref{eqn:lamb-form}, we cannot make use of the general reconstruction formulas. On the other hand, the angular momentum $\reJ$ and the world-sheet momentum $\pws$ can be found directly by integrating the Virasoro constraints. Using standard polar coordinates $Z = \cos\theta$ and $\Phi$ on the sphere, these constraints are given by
\be
  \frac{\tim{Z}^2+\spa{Z}^2}{1-Z^2} + (1-Z^2) (\tim{\Phi}^2+\spa{\Phi}^2) = 1
  \comma
  \frac{\tim{Z}\spa{Z}}{1-Z^2} + (1-Z^2) \, \tim{\Phi}\spa{\Phi} = 0 \; .
\ee
Aiming at the one-phase solution, one chooses the ansatz
\be \label{eqn:ansatz-Z-Phi}
  Z(\tau,\sigma) = z(k\sigma - \omega\tau)
  \comma
  \Phi(\tau,\sigma) = \varpi \tau + \varphi(k\sigma - \omega\tau)
\ee
for which one can solve the Virasoro constraints for the derivatives of $z$ and $\varphi$
\be \label{eqn:deri-z-varphi}
  z'^2       = \Bigbrk{\frac{k \varpi}{k^2-\omega^2}}^2 \, (z^2-z^2_{\mathrm{min}})(z^2_{\mathrm{max}}-z^2)
  \comma
  \varphi' = \frac{\omega \varpi}{k^2-\omega^2} \, \frac{z^2-z^2_{\mathrm{min}}}{1-z^2}
\ee
with
\be
  z_{\mathrm{min}} = \sqrt{1-\frac{1}{\varpi^2}}
  \comma
  z_{\mathrm{max}} = \sqrt{1-\frac{\omega^2}{k^2\varpi^2}}
  \; .
\ee
In order to see that this ansatz really corresponds to the kink train \eqref{eqn:SG-single-finite}, we have to compute the associated sine-Gordon field from the definition \eqref{eqn:defSGfield} which in polar coordinates reads
\be
  \cos 2\phi = \frac{\tim{Z}^2-\spa{Z}^2}{1-Z^2} + (1-Z^2)(\tim{\Phi}^2 - \spa{\Phi}^2) \; .
\ee
Using \eqref{eqn:ansatz-Z-Phi} and \eqref{eqn:deri-z-varphi}, this gives a relationship between $\phi$ and $z$ that otherwise only depends on the parameters $k$, $\omega$ and $\varpi$. This $\phi$ reduces to the kink train if and only if 
\be \label{eqn:AFZ-z}
  z(\tau,\sigma) = z_{\mathrm{max}} \, \JacobiDN\bigbrk{k\sigma-\omega\tau | m}
  \qquad\mbox{with}\qquad
  m = \frac{1}{k^2-\omega^2}
\ee
and
\be \label{eqn:cond-varpi}
  \varpi^2 = \frac{\omega^2 + (k^2-\omega^2)^2}{k^2} \; .
\ee

For the computation of $\reJ$ and $\pws$ one only needs \eqref{eqn:deri-z-varphi} and not the explicit solution \eqref{eqn:AFZ-z} because the integration over $\sigma$ can be substituted by an integration over $z$. For the angular momentum one finds
\be
  \reJ \eq \int d\sigma \: (1-z^2)\tim{\Phi}
       = 2 \int^{z_{\mathrm{max}}}_{z_{\mathrm{min}}} dz \: (1-z^2) \frac{\varpi-\omega \varphi'}{k \abs{z'}} \nln
       \eq \frac{2}{\sqrt{1+m^2\omega^2}} \bigbrk{\EllipticK(m)-\EllipticE(m)} \; .
\ee
The world-sheet momentum is computed from the separation of the endpoints and is given by
\be
  \pws \eq \Delta \Phi
           = 2\int^{z_{\mathrm{max}}}_{z_{\mathrm{min}}} dz\: \frac{\varphi'}{\abs{z'}} \nln
	     \eq 2\sqrt{1+m^2\omega^2} \Biggsbrk{  \frac{k}{\omega} \, \EllipticPi\biggbrk{ 1-\frac{k^2}{\omega^2} \bigg{|} m}
	                                         - \frac{\omega}{k} \, \EllipticK(m)} \; .
\ee
These expressions are valid in both the elementary region ($m<1$,$\varpi>1$) and the doubled region ($m>1$,$\varpi<1$).

\section{Semi-classical quantization and energy relations}
\label{sec:quantization}

In the previous sections we found for several classes of interacting two-magnon solutions the spatial period $L$, the temporal period $T$, the target-space energy $E$, and target-space angular momentum $J$ as exact functions of the parameters $k$ and $\omega$. Eliminating the parameters, we can, in principle, find\footnote{In the gauge chosen, we have $E(L,T)\equiv \frac{\sqrt{\lambda}}{2\pi}L$, see \eqref{eqn:target-space-energy}.}
\be \label{eqn:def-Delta}
  \Delta(L,T) := E(L,T) - J(L,T) \; ,
\ee
which is the relevant quantity for comparisons with SYM theory and which we will refer to as energy. This completely solves the \emph{classical} spectral problem, where $L$ and $T$ are continuous controllable parameters.

One would now like to convert \eqref{eqn:def-Delta} into a dispersion relation in order to make contact with previously known results (and possibly gain some insight into how integrability might work at finite size). This means that we need to replace $T$ by the (relative) magnon world-sheet momentum $p$ whose definition is, however, not unambiguous. In the decompactification limit the magnon momentum can be identified with the asymptotic angular separation of the string end points \cite{Hofman:2006xt}. In going to the two-magnon closed string solutions, where the individual solitons never cleanly separate, it is not clear how to define the magnon momentum geometrically. Instead we will make use of the fact that for the finite size case all the solutions are periodic in time, and so we can apply the Bohr-Sommerfeld condition to directly perform the semi-classical quantization. 

For a general system with periodic motion, period $T$, that is described by canonical variables $p$ and $q$, the method of Bohr-Sommerfeld quantization postulates the existence of an energy eigenstate whenever the condition
\be
\label{eqn:Bohr_Sommerfeld}
\int^T_0 dt \: p \, \tim{q} = 2\pi n \qquad\mbox{with}\quad n\in\Integers \; ,
\ee
is satisfied. Using the equations of motion, this condition can be cast into the differential form
\be \label{eqn:BS-quantization}
  T(\Delta) \, d\Delta = 2\pi \, dn \; ,
\ee
where $\Delta$ denotes the energy of the system. These semi-classical methods are also valid for field theories, see e.g.~\cite{Coleman:1975qj}. Applying \eqref{eqn:BS-quantization} to the finite-$J$ giant magnon solution and our two-magnon solutions gives us the excitation energy, $\Delta=E-J$, in terms of the action variable $n$.

For the magnon breathers this directly yields the energy formulas in terms of $n$ and we are able to straightforwardly find the finite size corrections to the infinite-$J$ results of \cite{Hofman:2006xt}. In addition, in the limit $J=0$ the fluxon breather simply becomes the circular pulsating string of \cite{Minahan:2002rc} and expanding in large $n$ we are able to match our expression for the energy with that previously found. For the scattering solution it remains to relate $n$ to the magnon momentum; an obvious candidate for a closed string is $\pws=\tfrac{2 \pi n}{L}$. However this relation corresponds to free excitations neither interacting with each other nor with boundaries and will be modified by the presence of a phase shift describing these interactions. We make use of the usual quantum mechanical relation, which can be extended to field theory \cite{Jackiw:1975im}, between the time delay and this phase. This can then be used to describe the scattering of solitons, a result which was used in the context of giant magnon scattering \cite{Hofman:2006xt}, and we will use analogous relations for our finite-$J$ two magnon states.

One obvious, but important point, is that the Bohr-Sommerfeld condition is only valid for large $n$ and does not include any zero-point energy. To include this effect one should use the WKB approximation as generalized by Dashen, Hasslacher and Neveu (DHM) \cite{Dashen:1974ci,Dashen:1974cj,Dashen:1974ck,Dashen:1975hd} to solitons in quantum field theories (this was done for the decompactified limit by \cite{Chen:2007vs}).

\subsection{Single magnon at finite $J$}

We apply our method to the single magnon case as a check and as an example of the general procedure. For nearly all our expansions we consider the solutions near the decompactification limit where the elliptic modulus controlling the spatial period approaches one i.e. $m_\sigma \simeq1\pm\epsilon$. We first expand the string length $L$, the period $T$, and the energy $\Delta=\tfrac{\sqrt{\lambda}}{2\pi}L-J$ in $\epsilon$ while keeping the velocity, or the analogous parameter for the breathers, fixed%
\footnote{We should note that this is merely a convenient intermediate step and one can equally well choose to fix some other parameter. 
In the end, we will express all our answers in terms of parameters with a gauge invariant target-space interpretation such as the angular momentum $J$ or $\Delta$.}.
We can then eliminate $\epsilon$ and express the period $T$ as a function of the energy $\Delta$ and the string length $L$, which then plays the role of a large expansion parameter. The integral of $T(\Delta,L)$ over $\Delta$ can then be related to the integer quantum number coming from the Bohr-Sommerfeld relation \eqref{eqn:BS-quantization}. In order to simplify subsequent formulas we introduce the rescaled energy
\be
  {\reen} = \frac{2\pi}{\sqrt{\lambda}}\Delta \; .
\ee

In the elementary region with $m=1-\epsilon$ and to order $\order(e^{-2L/\reen})$ we find for the period of the finite-$J$
giant magnon solution
\be
T \simeq \frac{-2 L}{\sqrt{4-{\reen}^2}}+ \left( \frac{4 L^2 {\reen} }{\sqrt{4-\reen^2}}-\frac{4L \reen^2 (2-\reen^2) }{(4-{\reen}^2)^{3/2}} \right) e^{ -\tfrac{2 L}{\reen}} \; .
\ee
According to \eqref{eqn:BS-quantization}, this should be equal to $\tfrac{(2\pi)^2}{\sqrt{\lambda}} \tfrac{dn}{d\reen}$ and so we must integrate this equation with respect to $\reen$. While It is not clear how to do this exactly it is straightforwardly done order by order in inverse powers of $L$,
\be \label{eqn:single-magnon-pws}
\frac{(2\pi)^2}{\sqrt{\lambda}} \frac{n}{L} = 2 \arcsin \frac{\reen}{2}+\left(\frac{2 \reen^3 }{\sqrt{4-\reen^2}}-\frac{4\reen^4 }{L \sqrt{4-\reen^2}}+\order\left(\frac{1}{L}\right)\right) e^{ -\tfrac{2 L}{\reen}}+\order\left(e^{ -\tfrac{4L}{\reen}}\right) \; .\nn\\
\ee
We see immediately that at leading order
\be
\Delta=\frac{\sqrt{\lambda}}{\pi}\sin \frac{1}{2}\left(\frac{2\pi n}{J+\Delta}\right) \; ,
\ee
where we used $L = \frac{2\pi}{\sqrt{\lambda}}(J+\Delta)$ and which agrees with the expected infinite magnon energy provided we identify $\pws=\tfrac{2\pi n}{J+\Delta}$. To compare the higher orders with the results of \cite{Arutyunov:2006gs} it is easiest to rewrite their dispersion relation as $\pws=\pws(\reen)$. Noting that $J=\tfrac{\sqrt{\lambda}}{2\pi}(L-\reen)$ and inverting their expression 
\be
\reen = 2 \sin \frac{\pws}{2}\left[1-4 \sin^2 \frac{\pws}{2} \, e^{-\left(\tfrac{L-\reen}{\sin\tfrac{\pws}{2}}+2\right)}\right]
\ee
we find 
\be
\pws(\reen) = 2 \arcsin\frac{\reen}{2}+\frac{ {2 \reen}^3}{ \sqrt{4-{\reen}^2}} \, e^{-\tfrac{2 L}{\reen }} \; .
\ee
If we set $\pws=\tfrac{2\pi n}{J+\Delta} = \frac{(2\pi)^2}{\sqrt{\lambda}} \frac{n}{L}$ and compare to \eqref{eqn:single-magnon-pws}, we see that we find agreement for the terms which are leading order in $L$ but not for the subleading terms. These corrections are presumably due to the fact that at this order for a finite magnon we must take into account the interactions between the excitation and the string endpoints. This is analogous to the fact that for two magnons the terms subleading in $L$  correspond to the interactions between magnons and hopefully will become clearer after the discussion of the magnon--anti-magnon scattering solution. Proceeding to the next order we find the same: at each order in $e^{-L/\reen}$ agreement in terms of leading order in $L$ given by
\be
\pws(\reen) = 2 \arcsin\frac{\reen}{2}+\frac{ {2 \reen}^3}{ \sqrt{4-{\reen}^2}} \, e^{-\tfrac{2 L}{\reen }} + 4 \reen L^3 \sqrt{4-\reen^2} \, e^{-\tfrac{4 L}{\reen }} \; ,
\ee
but disagreement at subleading orders. For comparison with later calculations let us record some of the higher order in exponential correction terms, but at each order again only keep the largest $L$ piece,
\be
\Delta \eq \frac{\sqrt{\lambda}}{\pi} \sin\frac{\pws}{2}\biggsbrk{1-4 \sin^2\frac{\pws}{2} e^{-\Leff}
\Bigbrk{
1
+2 \cos^2\frac{\pws}{2} \Leff^2 e^{-\Leff}
+8 \cos^4\frac{\pws}{2} \Leff^4 e^{-2\Leff} \nl
+\frac{128}{3}\cos^6\frac{\pws}{2} \Leff^6 e^{-3\Leff}
+\frac{800}{3}\cos^8\frac{\pws}{2} \Leff^8 e^{-4\Leff}
+\frac{9216}{5}\cos^{10}\frac{\pws}{2} \Leff^{10} e^{-5\Leff}+\ldots
} } \; , \nl
\ee
where we have introduced the effective length $\Leff=\tfrac{L}{\sin(\pws/2)}$.

In the doubled region, the finite-size corrections to the dispersion relation are very similar to those above but with the sign in front of the coefficient of the first correction flipped. We will not consider this case in detail here but it is worth keeping this in mind when we calculate the finite-size corrections to the two-magnon states in the doubled region. 

While we focus on the string theory near decompactification, it may also be interesting to consider the small radius limit. It is well known that the sine-Gordon theory simplifies dramatically in this ``UV'' regime essentially becoming the theory of a free scalar on a circle and it has proved useful to study the theory as a perturbation from this CFT, e.g. \cite{Zamolodchikov:1995xk}. For the string theory the underlying theory is of course already conformal however this symmetry can be spontaneously broken by expanding the gauge fixed theory about a non-trivial classical solution, for example near the BMN solution the string in light-cone gauge is described by a massive world-sheet theory. While we are not able to make any definite statements regarding the string theory in this limit as a small step in this direction we describe the classical energies near the zero length limit. 

For the single finite-$J$ magnon we take $k\sim\tfrac{1}{\epsilon}$ and $\omega\sim\tfrac{\beta}{\epsilon}$ which implies that the elliptic modulus is close to zero $m\sim\tfrac{\epsilon^2}{1-\beta^2}$. Thus we find
\be
L\sim\pi \epsilon  \quad\mbox{and}\quad  \pws\sim\pi(1-\beta)
\ee
so that
\be
\Delta \sim \frac{\sqrt{\lambda}L}{2\pi}\left(1-\frac{L}{2\pi (1-\beta^2)}\right) \; .
\ee
In this case $z_{\rm max}$ and $z_{\rm min}$ are both close to one and so the target-space extension of the string becomes very small. 
The angular momentum scales differently, it is of higher order, 
\be
J\sim\frac{\sqrt{\lambda}}{4 (1-\beta^2)}\pi \epsilon^2
\ee
and so to leading order the string has zero angular momentum. In this limit the string solution simplifies
considerably and it may be feasible to calculate quantum fluctuations about this background.

\subsection{Magnon breathers}

\subsubsection{Fluxonic magnon breather}
\label{sec:dispersion-fluxon-breather}

In the elementary region (i) we have $k^2+\omega^2 \le 1$ so that $m_\sigma\simeq1-\epsilon$, and in the doubled region (ii) we have $k^2+\omega^2 \ge 1$ so that $m_\sigma\simeq1+\epsilon$. For both of these cases we take $\omega^2\geq k(1-k)$ which is consistent with the decompactification limit. To $\order(e^{-4L/\reen})$ we find
\be
\label{eqn:fluxon_breather_period_exp}
  \frac{T}{2\pi} = \begin{cases}
      \frac{\reen}{\sqrt{\reen^2-16}} +\left(-\frac{256L}{(\reen^2-16)^{3/2}}+\frac{64\reen(-32+5\reen^2)}{(\reen^2-16)^{5/2}}\right) e^{-\tfrac{4L}{\reen}}& \mbox{for (i)} \; , \\
      \frac{\reen}{\sqrt{\reen^2-64}} +\left(\frac{1024L}{(\reen^2-64)^{3/2}}+\frac{256\reen(-64(-3+\ln 4)+\reen^2(-6+\ln 4))}{(\reen^2-64)^{5/2}}\right)e^{-\tfrac{4L}{\reen}}& \mbox{for (ii)} \; .
      \end{cases}
\ee

In the elementary region at leading order we have 
\be
2\pi dn= \frac{d\Delta}{ \sqrt{ 1-\frac{ 16}{ \reen^2} } }
\ee
and so $\Delta=\sqrt{n^2+\tfrac{4 \lambda}{\pi^2}}$ as in HM.

We can of course continue to the next order: 
\be
\label{eqn:fluxon_breather_energy_exp}
  \Delta= \begin{cases}
      \sqrt{n^2+\frac{4\lambda}{\pi^2}} +\left(\frac{16\lambda}{n^2\pi^2}\sqrt{n^2+\frac{4\lambda}{\pi^2}}-\frac{32\sqrt{\lambda}}{ n^2 L \pi}\left(n^2+\frac{4\lambda}{\pi^2}\right)\right)e^{-\Lelem} & \mbox{for (i)} \; , \\
         \sqrt{n^2+\frac{16\lambda}{\pi^2}} +\left(-\frac{64\lambda}{n^2\pi^2}\sqrt{n^2+\frac{16\lambda}{\pi^2}}+\frac{32\sqrt{\lambda}(5-\ln 4)}{n^2 L \pi}\left(n^2+\frac{16\lambda}{\pi^2}\right)\right)e^{-\Ldoub}& \mbox{for (ii)} \; ,
      \end{cases}\nn\\
\ee
where $\Lelem=4(J+\Delta)/\sqrt{n^2+\tfrac{4\lambda}{\pi^2}}$ and $\Ldoub=4(J+\Delta)/\sqrt{n^2+\tfrac{16\lambda}{\pi^2}}$.

As discussed in \secref{sec:reconstruction-fluxon-breather}, the fluxonic magnon breather for $J=0$ looks remarkably like the circular pulsating string of Minahan \cite{Minahan:2002rc} and indeed we expect that the two should be the same. Although the explicit solution for the circular string was not constructed in conformal gauge we can certainly match the target-space energies. Let us consider the pulsating circular string which is wrapped once around the sphere or in the notation of \cite{Minahan:2002rc}, $m=1$. The energy of the string is given by
\be \label{eqn:circ-puls-string}
 \Delta-4=2 n_M\left(1+\frac{\lambda}{4 (2n_M)^2}-\frac{1}{64}\left(\frac{\lambda}{ (2n_M)^2}\right)^2+\ldots\right).
\ee
In this formula the numerical constant on the left comes from initially treating the $n_M$ as finite and then the corrections are calculated assuming $n_M$, and consequently $\Delta$, is large. In order to compare we take in our  solution
$k=\epsilon$ which in turn implies that $\omega=1-\epsilon^2-\epsilon^4-2\epsilon^6$. As $J=0$ 
we have that
\be
\Delta=\frac{\sqrt{\lambda}}{2 \pi}L=\frac{\sqrt{\lambda}}{2 \epsilon}
\ee
and expressing the period as a function of the energy, $T=T(\Delta)$, we can integrate to find 
\be
\Delta=n+\frac{\lambda}{4 n}-\frac{\lambda^2}{64 n^3}+\ldots \; .
\ee
We see that this is the same as \eqref{eqn:circ-puls-string} if we identify $n=2 n_M$ and drop the constant from the left. Missing this ``ground state energy'' is the usual approximation made in Bohr-Sommerfeld quantization.

\subsubsection{Plasmonic magnon breather}

For the elementary region (i) where $k^2+\omega^2 \le 1$ and $m_\sigma=1+\epsilon$, and for the doubled region (ii) where $k^2+\omega^2 \ge 1$ and $m_\sigma=1-\epsilon$, we find to $\order(e^{-4L/\reen})$
\be
\label{eqn:plasmon_breather_period_exp}
  \frac{T}{2\pi} = \begin{cases}
      \frac{\reen}{\sqrt{\reen^2-16}} +\left(-\frac{256L}{(\reen^2-16)^{3/2}}+\frac{64\reen(-16(1+\ln 4)+\reen^2(4+\ln 4))}{(\reen^2-16)^{5/2}}\right) e^{-\tfrac{4L}{\reen}}& \mbox{for (i)} \; , \\
      \frac{\reen}{\sqrt{\reen^2-64}} +\left(\frac{1024L}{(\reen^2-64)^{3/2}}+\frac{256\reen(128-5 \reen^2 )}{(\reen^2-64)^{5/2}}\right)e^{-\tfrac{4L}{\reen}}& \mbox{for (ii)} \; ,
      \end{cases}
\ee
respectively. Integrating and inverting these equations gives
\be
\label{eqn:plasmon_breather_energy_exp}
  \Delta= \begin{cases}
      \sqrt{n^2+\frac{4\lambda}{\pi^2}} +\left(\frac{16\lambda}{n^2\pi^2}\sqrt{n^2+\frac{4\lambda}{\pi^2}}-\frac{8\sqrt{\lambda}(3+\ln 4)}{n^2 L \pi}\left(n^2+\frac{4\lambda}{\pi^2}\right)\right)e^{-\Lelem} & \mbox{for i)} \; , \\
         \sqrt{n^2+\frac{16\lambda}{\pi^2}} +\left(-\frac{64\lambda}{n^2\pi^2}\sqrt{n^2+\frac{16\lambda}{\pi^2}}+\frac{128\sqrt{\lambda}}{ n^2 L \pi}\left(n^2+\frac{16\lambda}{\pi^2}\right)\right)e^{-\Ldoub}& \mbox{for ii)} \; ,
      \end{cases}
\ee
where $\Lelem$ and $\Ldoub$ are the same as for the fluxonic magnon breather above.

\subsection{Magnon--anti-magnon scattering}

For the scattering solutions we first concentrate on the elementary region, $k^2-\omega^2 \ge 1$, with $m_{\sigma}=1-\epsilon$. To $\order(e^{-4L/\reen})$ we have
\be
\label{eqn:fluxon_osc_period_exp_ele}
  T\eq\frac{4L}{\sqrt{16-\reen^2}} +\frac{2\reen \ln \bigbrk{1-\tfrac{\reen^2}{16}}}{\sqrt{16-\reen^2}} \nl
    + \Biggsbrk{\frac{64L^2 \reen}{(16-\reen^2)^{3/2}}+\frac{32L \Bigbrk{3\reen^2(-8+\reen^2)+16(16-\reen^2)\ln\bigbrk{1-\tfrac{\reen^2}{16}}}}{(16-\reen^2)^{5/2}} \nl
    \hspace{10mm} -\frac{8\reen\Bigbrk{5\reen^4+16(32-5\reen^2)\ln \bigbrk{1-\tfrac{\reen^2}{16}}}}{(16-\reen^2)^{5/2}}} e^{-\tfrac{4L}{\reen}} \; .
\ee
Before continuing further let us consider the leading order and how it matches with previously known results. At leading order, expressing the energy and period in terms of  the 
velocity, $v=\frac{\omega}{k}$, we have 
\be
  \Delta=\frac{\sqrt{\lambda}}{\pi}\frac{2}{\gamma}
  \qquad\mbox{and}\qquad
  T= \frac{L}{v}+\frac{2}{v \gamma} \ln v^2 \; .
\ee 
The expression for the period consists of two terms which have the obvious interpretation as the time, $T_0$, for a freely moving particle of velocity $v$ to traverse the length of the string plus a correction, $T_{\rm delay}$, due to the interaction of the particle with a potential. Following the discussion of \cite{Jackiw:1975im} but applied to the string theory,  we interpret the center of mass motion of the two magnons as the motion of a particle of energy $\Delta$ and momentum $p$ (which being the relative momentum is twice the momentum $\pws$ of either of the individual magnons) moving in a periodic box of length $\tfrac{\sqrt{\lambda}}{2\pi}L$ with a potential. The boundary conditions imply
\be \label{eqn:split-p-delta}
 2\pi n = \tfrac{\sqrt{\lambda}}{2\pi} L \,  p + 2 \delta(\Delta) \; ,
\ee
where $2 \delta(\Delta)$ is the phase shift due to the interaction with the potential and corresponds to twice the phase that each magnon accrues on crossing the other. We now wish to find expressions for the momenta and phase  shift in terms of the energy, $\Delta$, and so we again make use of the Bohr-Sommerfeld rule 
\be
\label{eqn:BS-leading-order-scattering}
 T=T_0+ 2\, T_{\rm delay}=2\pi \frac{dn}{d\Delta} \; .
\ee
Substituting \eqref{eqn:split-p-delta} into \eqref{eqn:BS-leading-order-scattering} we identify the terms at 
each order in $L$ and thus get equations for $p$ and $\delta$. At leading order we
find 
\be
\frac{\sqrt{\lambda}}{2\pi}\frac{dp}{d\Delta}=\frac{1}{v}
\ee
so that, after using the relation between the velocity and the energy,
\be
 2 \pws = \int\frac{d{\reen}}{\sqrt{1-\tfrac{\reen^2}{16} } }
 \qquad\Rightarrow\qquad \Delta=2\frac{\sqrt{\lambda}}{\pi}\sin \frac{\pws}{2} \; ,
\ee
as we expect.
The terms at subleading order in $L$  give an expression for 
$\delta$,
\be
 \frac{\partial \delta(\Delta)}{\partial \Delta} = T_{\rm delay} = \frac{2}{v\gamma}\ln v \; .
\ee
This is exactly the result for the center of mass phase shift used by \cite{Hofman:2006xt} to calculate the scattering phase of two magnons in the infinite $J$ limit. Thus we not only reproduce the single magnon dispersion relation but furthermore we find the correction to the momenta from the AFS phase. With regard to our previous single magnon results we note that here, even in the decompactified limit, the subleading terms in $L$ correspond to corrections of the free dispersion relations. 
 
We now match our result for the finite size magnons with the spectral curve analysis of \cite{Minahan:2008re}.
We proceed as before: we have expanded the period in $e^{-4L/\reen}$ and  we further expand the coefficients of exponentials in powers of $L$ and then identify the leading term with the derivative of the momenta and the subleading terms with the phase shift. Thus now keeping the exponential finite-size correction and at the leading order in $L$ we have 
\be
 T=\frac{L}{\sqrt{1-\tfrac{\reen^2}{16}}}+\frac{L^2 {\reen}}{\left(1-\tfrac{\reen^2}{16}\right)^{3/2}} \, e^{-\tfrac{4L}{\reen}}.
\ee
which should, by equations \eqref{eqn:split-p-delta,eqn:BS-leading-order-scattering}, be equal to $\tfrac{\sqrt{\lambda}L}{2\pi}\tfrac{dp}{d\Delta}$. Integrating, and using $p=2 \pws$, we find 
\be
 \pws=2 \arcsin\frac{\reen}{4}+\frac{{\reen}^3}{8 (1-\tfrac{\reen^2}{16})^{3/2}} \, e^{-\tfrac{4L}{\reen}}
\ee
which can be inverted to give 
\be
 \Delta = 2 \frac{\sqrt{\lambda}}{\pi} \sin\frac{\pws}{2} \biggbrk{ 
 1 - 4 \frac{ \sin^2\frac{\pws}{2} }{ \cos^2\frac{\pws}{2} } \, e^{-\tfrac{L}{\sin\tfrac{\pws}{2}}} } \; .
\ee
Choosing the two magnon case and setting $p_1=\pws$ and $p_2=2\pi -\pws$ in the multi-magnon dispersion relation of  \cite{Minahan:2008re} we find perfect agreement.

We can repeat this to higher orders; expanding the period to order $e^{-24L/\reen}$ (i.e. sixth order) but again only keeping the leading term in $L$ at each order, we find
\be
T \eq \frac{L}{\sqrt{1-\tfrac{\reen^2}{16}}}
+\frac{L^2 {\reen}e^{-\tfrac{4L}{\reen}} }{\left(1-\tfrac{\reen^2}{16}\right)^{3/2}}
+\frac{4\times16 L^4 e^{-\tfrac{8L}{\reen}} }{{\reen} \left(1-\tfrac{\reen^2}{16}\right)^{3/2}}
+\frac{3 \times 16^3 L^6 e^{-\tfrac{12L}{\reen}} }{2{\reen}^3 \left(1-\tfrac{\reen^2}{16}\right)^{3/2}}
+\frac{ 2 \times16^5 L^8 e^{-\tfrac{16L}{\reen}} }{3 {\reen}^5 \left(1-\tfrac{\reen^2}{16}\right)^{3/2}}\nl
+\frac{ 5^3 \times 16^5 L^{10} e^{-\tfrac{20L}{\reen}} }{3{\reen}^7 \left(1-\tfrac{\reen^2}{16}\right)^{3/2}}
+\frac{ 6^3\times 16^7 L^{12} e^{-\tfrac{24L}{\reen}} }{ 5{\reen}^9 \left(1-\tfrac{\reen^2}{16}\right)^{3/2}}+\ldots \; .
\ee
This can in turn be integrated and inverted to find the dispersion relation
{\small
\be
 \Delta \eq 2 \frac{\sqrt{\lambda}}{\pi} \sin \frac{\pws}{2} \Biggsbrk{
1 
-4 \frac{ \sin^2  \frac{\pws}{2} }{\cos^2 \frac{\pws}{2} }e^{-\Leff}\biggsbrk{1
+2 \Leff^2 e^{-\Leff} 
+8 \Leff^4 e^{-2\Leff} 
+\frac{ 128}{3} \Leff^6 e^{-3\Leff}
 \nl
 \hspace{50mm}
+\frac{800}{3} \Leff^8 e^{-4\Leff} 
+\frac{ 9216}{5} \Leff^{10} e^{-5\Leff}+\ldots}} \; .
 \ee
}\noindent
with $\Leff=\tfrac{L}{\sin(\pws/2)}$. We note that expansion of the finite-size corrections involves the same coefficients as for the leading order single magnon. The expansion can be continued to yet higher orders and the obvious pattern seems to continue though what the resumed, closed  expression is remains undetermined.
 
We return to $\order(e^{-4L/\reen})$ and use the expressions for the period \eqref{eqn:fluxon_osc_period_exp_ele}, the Bohr-Sommerfeld relation \eqref{eqn:BS-leading-order-scattering}, and  boundary conditions \eqref{eqn:split-p-delta} to find the phase shift due to the interaction between magnons in a finite-volume. Expanding the 
coefficients of the exponential correction to the period we identify the derivative of the phase shift with the terms subleading in $L$. 
This can be integrated order by order in $L$ to give an expression for the first exponential finite size correction to the center of mass phase shift. 
We find the that at higher orders in exponential corrections the coefficients of the phase itself has contributions at different orders in $L$; though it is straightforward to keep subleading terms, for simplicity we keep only the leading order.
Thus we have
{\small
\be
\delta(\reen)\eq\frac{\sqrt{\lambda}}{\pi}\left(\sqrt{16-\reen^2}\left(2-\ln \left(1-\tfrac{\reen^2}{16}\right)\right)
+\frac{4\reen^2\left(-3\reen^2+16\ln \left(1-\tfrac{\reen^2}{16}\right)\right)}{(16-\reen^2)^{3/2}}e^{-\tfrac{4L}{\reen}}\right)\nl
+\order\left(\frac{1}{L}e^{-\tfrac{4L}{\reen}},e^{-\tfrac{8L}{\reen}}\right)
\ee}
or rewritten in terms of $\pws$ as
{\small
\be
\delta(\pws)\eq\frac{\sqrt{\lambda}}{\pi}\left(4\cos\tfrac{\pws}{2}  \left(1-\ln \left(\cos\tfrac{\pws}{2}\right)\right)-4 \left(6 \sin^2\tfrac{\pws}{2}-4\cos^2 \tfrac{\pws}{2} \ln \left(\cos\tfrac{\pws}{2}\right) \right) 
\frac{\sin^2 \tfrac{\pws}{2}}{ \cos^3 \tfrac{\pws}{2} }  e^{  -\Leff  } \right)
\nl
+\order \Big( \frac{1}{L} e^{-\Leff},e^{-2\Leff}\Big) \; .
\ee}
Even at leading order this is not the AFS phase evaluated at $p_1=\pws$ and $p_2=2\pi-\pws$, which is to be expected as we have performed the integration with respect to the center of mass energy and, as the result is not Lorentz invariant, calculating in different frames gives inequivalent answers. We currently do not know the string solution corresponding to two magnons moving with different velocities and finding the finite size corrections to the laboratory frame phase will have to be postponed until these solutions are known. 

\bigskip

We of course have corresponding results for the doubled region ($k^2-\omega^2 \le 1$, $m_{\sigma}=1+\epsilon$) which can be interpreted in a similar fashion. The period including the leading correction is similar to that of the elementary region:
{\small
\be
  T\eq
   \frac{8L }{\sqrt{64-\reen^2}} +\frac{4\reen \ln \left(1-\tfrac{\reen^2}{64}\right)}{\sqrt{64-\reen^2}}\nl
   - \Biggsbrk{\frac{128L^2 \reen}{(64-\reen^2)^{3/2}}-\frac{32L\Bigbrk{\reen^2(\reen^2(-7+\ln 4)+64(4-\ln 4))-128(64-\reen^2)\ln\bigbrk{1-\tfrac{\reen^2}{64}}}}{(64-\reen^2)^{5/2}}\nl
   +\frac{16 \reen\Bigbrk{\reen^4(-7+\ln16)+64\reen^2(2-\ln16)+64(\reen^2(6-\ln 4)-64(3-\ln 4))\ln \bigbrk{1-\tfrac{\reen^2}{64}}}}{(64-\reen^2)^{5/2}}}e^{-\tfrac{4L}{\reen}} \nn \\
\ee}
and this can be integrated so that
{\small
\be
\frac{(2\pi)^2}{\sqrt{\lambda}} \frac{n}{L} \eq 8 \arcsin \frac{\reen}{8}- 4\frac{\sqrt{64-\reen^2}}{L}\left(-2 +\ln \left(1-\frac{\reen^2}{64}\right)\right)\nl
-\Biggsbrk{ \frac{32 \reen^3}{(64-\reen^2)^{3/2}} + \frac{8\reen^2\Bigbrk{\reen^2(-7+\ln 4)+128 \ln  \bigbrk{1-\frac{\reen^2}{64}}}}{L(64-\reen^2)^{3/2}}\nl
+\frac{2\reen^3\Bigbrk{\reen^2(256(9-\ln 16)-\reen^2(21-5\ln 4))-128(64-\reen^2)(5-\ln 4) \ln  \bigbrk{1-\frac{\reen^2}{64}}}}{L^2(64-\reen^2)^{5/2}}}e^{-\tfrac{4L}{\reen}} \; . \nn \\
\ee}
Again we use \eqref{eqn:split-p-delta} but here identifying  $p=4\pws$ as we are in the doubled region and we can invert the above equation at leading order in $L$ to write
\be
\Delta \eq 4 \frac{\sqrt{\lambda}}{\pi} \sin \frac{\pws}{2} \left(1 +4\frac{\sin^2 \frac{\pws}{2}}{\cos^2 \frac{\pws}{2}} \, e^{-\tfrac{L}{2 \sin \tfrac{\pws}{2}}}\right) \; .
\ee
We note that this is four times the single magnon dispersion relation but with the sign of the exponential corrections flipped. This is consistent with the string being in the doubled region and the state consisting of two ``helical'' strings.

\section{Conclusions and outlook}

As a step toward understanding the AdS/CFT duality, and particularly the role of integrability, for states of finite $R$-charge we have studied the classical finite-volume bosonic string moving on $\Reals\times\Sphere^2$. Making use of the connection between the $\grO(3)$ sigma-model and sine-Gordon theory we have found explicit two-phase solutions to the string equations of motion with periodic boundary conditions. We start by considering the known periodic solutions of sine-Gordon, the fluxon oscillation, the fluxon breather and the plasmon breather, and reconstruct the corresponding string solutions. The inverse map is non-local and therefore it is very non-trivial to find these string states. Fortunately the classical relations between surfaces of constant curvature and sine-Gordon theory provide a convenient formalism for the string reconstruction. This allows us to integrate the equations and to find the target space string for the two-phase solutions corresponding to solitons in the center of mass. Additionally, we compute the periods, the target-space energy and the angular momentum for these string configurations.

The two-phase solutions turn out to be significantly simpler than one would naively expect. They are given in terms of elliptic functions rather than the more general hyper-elliptic functions which generically correspond to the two-cut Riemann surfaces that follow from the algebraic curve analysis. This simplicity is a consequence of the string solutions having vanishing total world-sheet momentum which guarantees that they are indeed physical closed strings satisfying the world-sheet constraints. 

In the context of integrability it has proved useful to admit unphysical strings which serve as building blocks for physical ones and, if one aims at generalizing the asymptotic Bethe equations to also describe the finite size spectrum, it would be important to find periodic string solutions for magnons with different velocities. Since the gauge fixed world-sheet theory is no longer Lorentz invariant this is unfortunately not simply a matter of performing a boost. Another path, but perhaps just as complicated, would be to construct the three magnon solutions. There are explicit formulas in terms of Riemann theta functions for sine-Gordon three-phase solutions \cite{Bobenko:1984} though at this point one may as well use the generic string solutions constructed in \cite{Dorey:2006zj,Dorey:2006mx}. The three-magnon result would of course also be interesting as it may shine light on the question of what are the useful quantities to generalize to arbitrary magnon states. The three phase solutions would in addition to the spatial and temporal periods have a third ``period'' in an independent combination of the space and time coordinates. This new period would presumably correspond after quantization to the second independent excitation number describing a three-magnon state. 

Having the two-magnon periodic solutions in hand we calculate the finite size corrections to their dispersion relations. All the solutions are temporally periodic so we use, as a first approximation, the Bohr-Sommerfeld condition to relate their energies to the quantum oscillation number. As mentioned earlier it would be interesting to carry out a proper WKB analysis of these solitonic solutions \`{a} la DHN which would correctly account for the zero-point contributions (within the context of sine-Gordon theory this was carried out by \cite{Mussardo:2004zn}). For the string states corresponding to the breathers the relation between the quantum number and the energy is straightforward and we directly find the exponentially suppressed corrections. While the breather solutions describe bound states of two solitons in sine-Gordon theory, results in the decompactification limit \cite{Dorey:2007xn} suggest that the magnon breathers are presumably not actual bound states but rather superpositions of BPS magnons with opposite charges. It would be interesting to check that this is indeed the case. This would require studying finite-$J$ solutions on $\Reals\times\Sphere^3$ which may be possible using methods similar to those discussed here and making use of the relation between the $\grO(4)$ sigma model and complex sine-Gordon. Additionally we are able to make contact with the circular pulsating string where, as the angular momentum $J$ vanishes, the effective length $L/D$ becomes unity and hence the exponential corrections are all of order one. Here we expand in large $n$ and we find agreement with the expression previously calculated by Minahan.

For the scattering states the relation is slightly more complicated and we break the answer into two parts: the dispersion relation in terms of the magnon momentum and the phase due to magnon interactions. This splitting is somewhat arbitrary as there is no regime where the individual magnons are insensitive to each other. This fact is immediately apparent from previous calculations of the multi-magnon dispersion relation where the energy of each magnon depends on the momenta of all the others and indeed the energy of the two-magnon state is not simply the sum of two individual magnon energies. Nonetheless, by making use of the different dependences on the size of the system we label the different contributions. The leading part in the string length $L$ at each order in $e^{-L}$ is considered as the term giving rise to the dispersion relation and the remaining, sub-leading terms as corresponding to the phase shift due to interactions. This allows us to make contact with previous calculations and at leading order in $L$ we do indeed find agreement with the dispersion relation of the finite-size single magnon of AFZ and the multi-magnon dispersion relation of Minahan and Ohlsson Sax. We are further able to calculate the leading order terms to higher orders in $e^{-L}$ and find a somewhat regular pattern. It would of course be interesting to find closed all order expressions for the magnon energies even if only at the classical level. Keeping terms at sub-leading powers in $L$ we find the finite size corrections to the analogue of the scattering phase. However, as mentioned above, from the two-phase solutions we can calculate this phase only in the center of mass frame and it is not clear how to find the analogous result for arbitrary momenta.

\bigskip
\subsection*{Acknowledgments}
\bigskip

The authors gladly acknowledge useful conversations with Diego Hofman, Joe Minahan, Thomas Neukirchner and Peter Orland. We would also like to thank the Isaac Newton Institute in Cambridge and the organizers of the SIS workshop, where this work was initiated, for their kind hospitality.

\appendix

\section{Elliptic functions} \label{app:elliptic}

We use the conventions of Abramowitz and Stegun \cite{abramowitz+stegun}. The elliptic integrals of first, second and third kind are defined, respectively, as
\begin{align}
  \EllipticF(\varphi|m) & := \int_0^\varphi \frac{d\theta}{\sqrt{1-m\sin^2\theta}} \; , \label{eqn:ellF} \\
  \EllipticE(\varphi|m) & := \int_0^\varphi \sqrt{1-m\sin^2\theta} \, d\theta \; , \label{eqn:ellE} \\
  \EllipticPi(n,\varphi|m) & := \int_0^\varphi \frac{d\theta}{(1-n\sin^2\theta)\sqrt{1-m\sin^2\theta}} \; . \label{eqn:ellPi}
\end{align}
The complete elliptic integrals are denoted by
\be
  \EllipticK(m) = \EllipticF(\tfrac{\pi}{2}|m)
  \comma
  \EllipticE(m) = \EllipticE(\tfrac{\pi}{2}|m)
  \comma
  \EllipticPi(n|m) = \EllipticPi(n,\tfrac{\pi}{2}|m)
  \; ,
\ee
and $\EllipticK'(m) = \EllipticK(1-m)$. The elliptic amplitude $\am$ is defined as the inverse of $\EllipticF$
\be
  \varphi(u) = \am(u|m)  \quad\Leftrightarrow\quad  u(\varphi) = \EllipticF(\varphi|m) \; .
\ee
Periodicity for $r,s\in \Integers$
\be
  \am(u+2r\EllipticK(m)+2is\EllipticK'(m)|m) = \am(u|m) + r \pi
\ee
Define also Jacobi elliptic functions
\begin{align}
  \JacobiSN(u|m) & = \sin\am(u|m) \\
  \JacobiCN(u|m) & = \cos\am(u|m) \\
  \JacobiDN(u|m) & = \sqrt{1-m \JacobiSN^2(u|m)}
\end{align}
\be
  \mathrm{pq}(u|m) = \frac{\mathrm{pr}(u|m)}{\mathrm{qr}(u|m)}
  \comma
  \mathrm{pp}(u|m) = 1
\ee
where $\mathrm{p}$, $\mathrm{q}$ and $\mathrm{r}$ are any of the letters $\mathrm{s}$, $\mathrm{c}$, $\mathrm{d}$ and $\mathrm{n}$.

Some useful identities that we applied are
\begin{align}
  & \EllipticF(i\arcsinh\JacobiSC(u|m) | 1-m) = i u \; , \\
  & \EllipticPi(n,i\arcsinh\tan z | 1-m) = \frac{i}{1-n} \bigsbrk{ \EllipticF(z|m) - n \EllipticPi(1-n,z|m) } \; ,
\end{align}
and
\begin{align}
  & \EllipticF(z|m)    = \frac{1}{\sqrt{m}} \EllipticF\lrbrk{\arcsin(\sqrt{m}\sin z)\big{|}\tfrac{1}{m}} \; , \\
  & \EllipticPi(n,z|m) = \frac{1}{\sqrt{m}} \EllipticPi\lrbrk{\tfrac{n}{m},\arcsin(\sqrt{m}\sin z)\big{|}\tfrac{1}{m}} \; .
\end{align}

\bibliographystyle{nb}
\bibliography{finitemagnons}

\begin{thebibliography}{10}
\ifx\href\asklfhas\newcommand{\href}[2]{#2}\fi
\raggedright
\small
\parskip 0pt

\bibitem{Maldacena:1997re}
J.~M.~Maldacena,
\textit{``The large N limit of superconformal field theories and
  supergravity''},
\textsf{Adv.~Theor.~Math.~Phys.~2,~231~(1998)},
\href{http://arXiv.org/abs/hep-th/9711200}{\texttt{hep-th/9711200}}.
%
\bibitem{Witten:1998qj}
E.~Witten,
\textit{``Anti-de Sitter space and holography''},
\textsf{Adv.~Theor.~Math.~Phys.~2,~253~(1998)},
\href{http://arXiv.org/abs/hep-th/9802150}{\texttt{hep-th/9802150}}.
%
\bibitem{Gubser:1998bc}
S.~S.~Gubser, I.~R.~Klebanov and A.~M.~Polyakov,
\textit{``Gauge theory correlators from non-critical string theory''},
\textsf{Phys.~Lett.~B428,~105~(1998)},
\href{http://arXiv.org/abs/hep-th/9802109}{\texttt{hep-th/9802109}}.
%
\bibitem{Minahan:2002ve}
J.~A.~Minahan and K.~Zarembo,
\textit{``The Bethe-ansatz for $\mathcal{N}=4$ super Yang-Mills''},
\textsf{JHEP~0303,~013~(2003)},
\href{http://arXiv.org/abs/hep-th/0212208}{\texttt{hep-th/0212208}}.
%
\bibitem{Beisert:2003yb}
N.~Beisert and M.~Staudacher,
\textit{``The $\mathcal{N}=4$ SYM Integrable Super Spin Chain''},
\textsf{Nucl.~Phys.~B670,~439~(2003)},
\href{http://arXiv.org/abs/hep-th/0307042}{\texttt{hep-th/0307042}}.
%
\bibitem{Beisert:2003tq}
N.~Beisert, C.~Kristjansen and M.~Staudacher,
\textit{``The dilatation operator of $\mathcal{N}=4$ conformal super Yang-Mills
  theory''},
\textsf{Nucl.~Phys.~B664,~131~(2003)},
\href{http://arXiv.org/abs/hep-th/0303060}{\texttt{hep-th/0303060}}.
%
\bibitem{Mandal:2002fs}
G.~Mandal, N.~V.~Suryanarayana and S.~R.~Wadia,
\textit{``Aspects of semiclassical strings in $AdS_5$''},
\textsf{Phys.~Lett.~B543,~81~(2002)},
\href{http://arXiv.org/abs/hep-th/0206103}{\texttt{hep-th/0206103}}.
%
\bibitem{Bena:2003wd}
I.~Bena, J.~Polchinski and R.~Roiban,
\textit{``Hidden symmetries of the $AdS_5\times S^5$ superstring''},
\textsf{Phys.~Rev.~D69,~046002~(2004)},
\href{http://arXiv.org/abs/hep-th/0305116}{\texttt{hep-th/0305116}}.
%
\bibitem{Berenstein:2002jq}
D.~E.~Berenstein, J.~M.~Maldacena and H.~S.~Nastase,
\textit{``{Strings in flat space and pp waves from N = 4 super Yang Mills}''},
\textsf{JHEP~0204,~013~(2002)},
\href{http://arXiv.org/abs/hep-th/0202021}{\texttt{hep-th/0202021}}.
%
\bibitem{Beisert:2005tm}
N.~Beisert,
\textit{``The $su(2|2)$ dynamic S-matrix''},
\href{http://arXiv.org/abs/hep-th/0511082}{\texttt{hep-th/0511082}}.
%
\bibitem{Staudacher:2004tk}
M.~Staudacher,
\textit{``The factorized S-matrix of CFT/AdS''},
\textsf{JHEP~0505,~054~(2005)},
\href{http://arXiv.org/abs/hep-th/0412188}{\texttt{hep-th/0412188}}.
%
\bibitem{Beisert:2006ib}
N.~Beisert, R.~Hernandez and E.~Lopez,
\textit{``A crossing-symmetric phase for $AdS_5 \times S^5$ strings''},
\textsf{JHEP~0611,~070~(2006)},
\href{http://arXiv.org/abs/hep-th/0609044}{\texttt{hep-th/0609044}}.
%
\bibitem{Arutyunov:2004vx}
G.~Arutyunov, S.~Frolov and M.~Staudacher,
\textit{``Bethe ansatz for quantum strings''},
\textsf{JHEP~0410,~016~(2004)},
\href{http://arXiv.org/abs/hep-th/0406256}{\texttt{hep-th/0406256}}.
%
\bibitem{Hernandez:2006tk}
R.~Hernandez and E.~Lopez,
\textit{``Quantum corrections to the string Bethe ansatz''},
\textsf{JHEP~0607,~004~(2006)},
\href{http://arXiv.org/abs/hep-th/0603204}{\texttt{hep-th/0603204}}.
%
\bibitem{Gromov:2007cd}
N.~Gromov and P.~Vieira,
\textit{``Constructing the AdS/CFT dressing factor''},
\href{http://arXiv.org/abs/hep-th/0703266}{\texttt{hep-th/0703266}}.
%
\bibitem{Janik:2006dc}
R.~A.~Janik,
\textit{``The $AdS_5 \times S^5$ superstring worldsheet S-matrix and crossing
  symmetry''},
\textsf{Phys.~Rev.~D73,~086006~(2006)},
\href{http://arXiv.org/abs/hep-th/0603038}{\texttt{hep-th/0603038}}.
%
\bibitem{Beisert:2006ez}
N.~Beisert, B.~Eden and M.~Staudacher,
\textit{``Transcendentality and crossing''},
\textsf{J.~Stat.~Mech.~0701,~P021~(2007)},
\href{http://arXiv.org/abs/hep-th/0610251}{\texttt{hep-th/0610251}}.
%
\bibitem{Bern:2006ew}
Z.~Bern, M.~Czakon, L.~J.~Dixon, D.~A.~Kosower and V.~A.~Smirnov,
\textit{``The Four-Loop Planar Amplitude and Cusp Anomalous Dimension in
  Maximally Supersymmetric Yang-Mills Theory''},
\textsf{Phys.~Rev.~D75,~085010~(2007)},
\href{http://arXiv.org/abs/hep-th/0610248}{\texttt{hep-th/0610248}}.
%
\bibitem{Cachazo:2006az}
F.~Cachazo, M.~Spradlin and A.~Volovich,
\textit{``Four-Loop Cusp Anomalous Dimension From Obstructions''},
\textsf{Phys.~Rev.~D75,~105011~(2007)},
\href{http://arXiv.org/abs/hep-th/0612309}{\texttt{hep-th/0612309}}.
%
\bibitem{Beisert:2007hz}
N.~Beisert, T.~McLoughlin and R.~Roiban,
\textit{``The Four-Loop Dressing Phase of $\mathcal{N} = 4$ SYM''},
\href{http://arXiv.org/abs/arXiv:0705.0321 [hep-th]}{\texttt{arXiv:0705.0321
  [hep-th]}}.
%
\bibitem{Beisert:2005fw}
N.~Beisert and M.~Staudacher,
\textit{``Long-range $PSU(2,2|4)$ Bethe ansaetze for gauge theory and
  strings''},
\textsf{Nucl.~Phys.~B727,~1~(2005)},
\href{http://arXiv.org/abs/hep-th/0504190}{\texttt{hep-th/0504190}}.
%
\bibitem{Ambjorn:2005wa}
J.~Ambjorn, R.~A.~Janik and C.~Kristjansen,
\textit{``{Wrapping interactions and a new source of corrections to the
  spin-chain / string duality}''},
\textsf{Nucl.~Phys.~B736,~288~(2006)},
\href{http://arXiv.org/abs/hep-th/0510171}{\texttt{hep-th/0510171}}.
%
\bibitem{Kotikov:2007cy}
A.~V.~Kotikov, L.~N.~Lipatov, A.~Rej, M.~Staudacher and V.~N.~Velizhanin,
\textit{``Dressing and Wrapping''},
\href{http://arXiv.org/abs/arXiv:0704.3586 [hep-th]}{\texttt{arXiv:0704.3586
  [hep-th]}}.
%
\bibitem{Keeler:2008ce}
C.~A.~Keeler and N.~Mann,
\textit{``{Wrapping Interactions and the Konishi Operator}''},
\href{http://arXiv.org/abs/arXiv:0801.1661 [hep-th]}{\texttt{arXiv:0801.1661
  [hep-th]}}.
%
\bibitem{Fiamberti:2007rj}
F.~Fiamberti, A.~Santambrogio, C.~Sieg and D.~Zanon,
\textit{``{Wrapping at four loops in N=4 SYM}''},
\href{http://arXiv.org/abs/arXiv:0712.3522 [hep-th]}{\texttt{arXiv:0712.3522
  [hep-th]}}.
%
\bibitem{Sieg:2005kd}
C.~Sieg and A.~Torrielli,
\textit{``{Wrapping interactions and the genus expansion of the 2- point
  function of composite operators}''},
\textsf{Nucl.~Phys.~B723,~3~(2005)},
\href{http://arXiv.org/abs/hep-th/0505071}{\texttt{hep-th/0505071}}.
%
\bibitem{Eden:2007rd}
B.~Eden,
\textit{``{Boxing with Konishi}''},
\href{http://arXiv.org/abs/arXiv:0712.3513 [hep-th]}{\texttt{arXiv:0712.3513
  [hep-th]}}.
%
\bibitem{Klose:2006zd}
T.~Klose, T.~McLoughlin, R.~Roiban and K.~Zarembo,
\textit{``Worldsheet scattering in $AdS_5 \times S^5$''},
\textsf{JHEP~0703,~094~(2007)},
\href{http://arXiv.org/abs/hep-th/0611169}{\texttt{hep-th/0611169}}.
%
\bibitem{Maldacena:2006rv}
J.~Maldacena and I.~Swanson,
\textit{``Connecting giant magnons to the pp-wave: An interpolating limit of
  $AdS_5 \times S^5$''},
\href{http://arXiv.org/abs/hep-th/0612079}{\texttt{hep-th/0612079}}.
%
\bibitem{Klose:2007wq}
T.~Klose and K.~Zarembo,
\textit{``Reduced sigma-model on $AdS_5 \times S^5$: one-loop scattering
  amplitudes''},
\textsf{JHEP~0702,~071~(2007)},
\href{http://arXiv.org/abs/hep-th/0701240}{\texttt{hep-th/0701240}}.
%
\bibitem{Klose:2007rz}
T.~Klose, T.~McLoughlin, J.~A.~Minahan and K.~Zarembo,
\textit{``World-sheet scattering in $AdS_5 \times S^5$ at two loops''},
\href{http://arXiv.org/abs/arXiv:0704.3891 [hep-th]}{\texttt{arXiv:0704.3891
  [hep-th]}}.
%
\bibitem{Arutyunov:2006yd}
G.~Arutyunov, S.~Frolov and M.~Zamaklar,
\textit{``The Zamolodchikov-Faddeev algebra for $AdS_5 \times S^5$
  superstring''},
\textsf{JHEP~0704,~002~(2007)},
\href{http://arXiv.org/abs/hep-th/0612229}{\texttt{hep-th/0612229}}.
%
\bibitem{Hofman:2006xt}
D.~M.~Hofman and J.~M.~Maldacena,
\textit{``Giant magnons''},
\textsf{J.~Phys.~A39,~13095~(2006)},
\href{http://arXiv.org/abs/hep-th/0604135}{\texttt{hep-th/0604135}}.
%
\bibitem{Schafer-Nameki:2005tn}
S.~Schafer-Nameki, M.~Zamaklar and K.~Zarembo,
\textit{``Quantum corrections to spinning strings in $AdS_5 \times S^5$ and
  Bethe ansatz: A comparative study''},
\textsf{JHEP~0509,~051~(2005)},
\href{http://arXiv.org/abs/hep-th/0507189}{\texttt{hep-th/0507189}}.
%
\bibitem{Schafer-Nameki:2006ey}
S.~Schafer-Nameki, M.~Zamaklar and K.~Zarembo,
\textit{``How accurate is the quantum string Bethe ansatz?''},
\textsf{JHEP~0612,~020~(2006)},
\href{http://arXiv.org/abs/hep-th/0610250}{\texttt{hep-th/0610250}}.
%
\bibitem{Arutyunov:2006gs}
G.~Arutyunov, S.~Frolov and M.~Zamaklar,
\textit{``Finite-size effects from giant magnons''},
\textsf{Nucl.~Phys.~B778,~1~(2007)},
\href{http://arXiv.org/abs/hep-th/0606126}{\texttt{hep-th/0606126}}.
%
\bibitem{Astolfi:2007uz}
D.~Astolfi, V.~Forini, G.~Grignani and G.~W.~Semenoff,
\textit{``{Gauge invariant finite size spectrum of the giant magnon}''},
\textsf{Phys.~Lett.~B651,~329~(2007)},
\href{http://arXiv.org/abs/hep-th/0702043}{\texttt{hep-th/0702043}}.
%
\bibitem{Dorey:2006dq}
N.~Dorey,
\textit{``{Magnon bound states and the AdS/CFT correspondence}''},
\textsf{J.~Phys.~A39,~13119~(2006)},
\href{http://arXiv.org/abs/hep-th/0604175}{\texttt{hep-th/0604175}}.
%
\bibitem{Okamura:2006zv}
K.~Okamura and R.~Suzuki,
\textit{``{A perspective on classical strings from complex sine- Gordon
  solitons}''},
\textsf{Phys.~Rev.~D75,~046001~(2007)},
\href{http://arXiv.org/abs/hep-th/0609026}{\texttt{hep-th/0609026}}.
%
\bibitem{Hatsuda:2008gd}
Y.~Hatsuda and R.~Suzuki,
\textit{``{Finite-Size Effects for Dyonic Giant Magnons}''},
\href{http://arXiv.org/abs/arXiv:0801.0747 [hep-th]}{\texttt{arXiv:0801.0747
  [hep-th]}}.
%
\bibitem{Luscher:1985dn}
M.~L{\"u}scher,
\textit{``{Volume Dependence of the Energy Spectrum in Massive Quantum Field
  Theories. 1. Stable Particle States}''},
\textsf{Commun.~Math.~Phys.~104,~177~(1986)}.
%
\bibitem{Janik:2007wt}
R.~A.~Janik and T.~Lukowski,
\textit{``{Wrapping interactions at strong coupling -- the giant magnon}''},
\textsf{Phys.~Rev.~D76,~126008~(2007)},
\href{http://arXiv.org/abs/arXiv:0708.2208 [hep-th]}{\texttt{arXiv:0708.2208
  [hep-th]}}.
%
\bibitem{Heller:2008at}
M.~P.~Heller, R.~A.~Janik and T.~Lukowski,
\textit{``{A new derivation of L{\"u}scher F-term and fluctuations around the
  giant magnon}''},
\href{http://arXiv.org/abs/arXiv:0801.4463 [hep-th]}{\texttt{arXiv:0801.4463
  [hep-th]}}.
%
\bibitem{Gromov:2008ie}
N.~Gromov, S.~Schafer-Nameki and P.~Vieira,
\textit{``{Quantum Wrapped Giant Magnon}''},
\href{http://arXiv.org/abs/arXiv:0801.3671 [hep-th]}{\texttt{arXiv:0801.3671
  [hep-th]}}.
%
\bibitem{Arutyunov:2007tc}
G.~Arutyunov and S.~Frolov,
\textit{``{On String S-matrix, Bound States and TBA}''},
\textsf{JHEP~0712,~024~(2007)},
\href{http://arXiv.org/abs/arXiv:0710.1568 [hep-th]}{\texttt{arXiv:0710.1568
  [hep-th]}}.
%
\bibitem{Minahan:2008re}
J.~A.~Minahan and O.~Ohlsson~Sax,
\textit{``Finite size effects for giant magnons on physical strings''},
\href{http://arXiv.org/abs/arXiv:0801.2064 [hep-th]}{\texttt{arXiv:0801.2064
  [hep-th]}}.
%
\bibitem{Dorey:2006zj}
N.~Dorey and B.~Vicedo,
\textit{``{On the dynamics of finite-gap solutions in classical string
  theory}''},
\textsf{JHEP~0607,~014~(2006)},
\href{http://arXiv.org/abs/hep-th/0601194}{\texttt{hep-th/0601194}}.
%
\bibitem{Pohlmeyer:1975nb}
K.~Pohlmeyer,
\textit{``Integrable Hamiltonian Systems and Interactions Through Quadratic
  Constraints''},
\textsf{Commun.~Math.~Phys.~46,~207~(1976)}.
%
\bibitem{Mikhailov:2005qv}
A.~Mikhailov,
\textit{``An action variable of the sine-Gordon model''},
\textsf{J.~Geom.~Phys.~56,~2429~(2006)},
\href{http://arXiv.org/abs/hep-th/0504035}{\texttt{hep-th/0504035}}.
%
\bibitem{Mikhailov:2007xr}
A.~Mikhailov and S.~Schafer-Nameki,
\textit{``{Sine-Gordon-like action for the Superstring in
  $\AdS_5\times\Sphere^5$}''},
\href{http://arXiv.org/abs/arXiv:0711.0195 [hep-th]}{\texttt{arXiv:0711.0195
  [hep-th]}}.
%
\bibitem{Grigoriev:2007bu}
M.~Grigoriev and A.~A.~Tseytlin,
\textit{``{Pohlmeyer reduction of $\AdS_5 \times \Sphere^5$ superstring sigma
  model}''},
\href{http://arXiv.org/abs/arXiv:0711.0155 [hep-th]}{\texttt{arXiv:0711.0155
  [hep-th]}}.
%
\bibitem{Forest:1982ji}
M.~G.~Forest and D.~W.~Mclaughlin,
\textit{``Spectral theory for the periodic sine-Gordon equation: a concrete
  viewpoint''},
\textsf{J.~Math.~Phys.~23,~1248~(1982)}.
%
\bibitem{Costabile:1978}
G.~Costabile, R.~D.~Parmentier, B.~Savo, D.~W.~McLaughlin and A.~C.~Scott,
\textit{``Exact solutions of the sine-Gordon equation describing oscillations
  in a long (but finite) Josephson junction''},
\textsf{Appl.~Phys.~Lett.~32(9),~587~(1978)}.
%
\bibitem{Lamb:1971zz}
G.~L.~Lamb,
\textit{``{Analytical Descriptions of Ultrashort Optical Pulse Propagation in a
  Resonant Medium}''},
\textsf{Rev.~Mod.~Phys.~43,~99~(1971)}.
%
\bibitem{Gu:1996zm}
H.~Hesheng,
\textit{``{Soliton and differential geometry}''},
in: \textit{``Soliton theory and its applications''},
ed.: C.-H.~Gu,
Berlin, Germany: Springer (1995),
403p.
%
\bibitem{Dorey:2007xn}
N.~Dorey, D.~M.~Hofman and J.~Maldacena,
\textit{``On the singularities of the magnon S-matrix''},
\href{http://arXiv.org/abs/hep-th/0703104}{\texttt{hep-th/0703104}}.
%
\bibitem{Spradlin:2006wk}
M.~Spradlin and A.~Volovich,
\textit{``Dressing the giant magnon''},
\textsf{JHEP~0610,~012~(2006)},
\href{http://arXiv.org/abs/hep-th/0607009}{\texttt{hep-th/0607009}}.
%
\bibitem{Minahan:2002rc}
J.~A.~Minahan,
\textit{``{Circular semiclassical string solutions on $\AdS_5\times
  \Sphere^5$}''},
\textsf{Nucl.~Phys.~B648,~203~(2003)},
\href{http://arXiv.org/abs/hep-th/0209047}{\texttt{hep-th/0209047}}.
%
\bibitem{Metsaev:1998it}
R.~R.~Metsaev and A.~A.~Tseytlin,
\textit{``Type IIB superstring action in $AdS_5 \times S^5$ background''},
\textsf{Nucl.~Phys.~B533,~109~(1998)},
\href{http://arXiv.org/abs/hep-th/9805028}{\texttt{hep-th/9805028}}.
%
\bibitem{Lund:1976ze}
F.~Lund and T.~Regge,
\textit{``{Unified Approach to Strings and Vortices with Soliton Solutions}''},
\textsf{Phys.~Rev.~D14,~1524~(1976)}.
%
\bibitem{Lund:1976xd}
F.~Lund,
\textit{``{Note on the Geometry of the Nonlinear Sigma Model in Two-
  Dimensions}''},
\textsf{Phys.~Rev.~D15,~1540~(1977)}.
%
\bibitem{Matveev:1976mj}
V.~B.~Matveev,
\textit{``{Abelian Functions and Solitons}''},
Univ. of Wroclaw Report No. 373 (unpublished).
%
\bibitem{Kozel:1976}
V.~A.~Kozel and V.~P.~Kotlyarov,
\textit{``{Almost periodical solutions of the equation $u_{tt}-u_{xx}+\sin
  u=0$}''},
\textsf{Dokl.~Akad.~Nauk~Ukr.~SSR,~Ser.~A~10,~878~(1976)}.
%
\bibitem{Fulton:1977}
T.~A.~Fulton,
\textit{``{Equivalent circuits and analogs of the Josephson effect}''},
in: \textit{``Superconductor and applications. squids and machines''},
ed.: B.~B.~Schwartz and S.~Foner,
Plenum Press, New York (1977),
125-187p.
%
\bibitem{Chebyshev:1946}
P.~L.~Chebyshev,
\textit{``{{\"U}ber den Schnitt von Bekleidungen}''},
\textsf{Usp.~Mat.~Nauk~1,~38~(1946)}.
%
\bibitem{Coleman:1975qj}
S.~R.~Coleman,
\textit{``Classical Lumps and their Quantum Descendents''},
Lectures delivered at Int. School of Subnuclear Physics, Ettore Majorana,
  Erice, Sicily, Jul 11-31, 1975.
%
\bibitem{Jackiw:1975im}
R.~Jackiw and G.~Woo,
\textit{``{Semiclassical Scattering of Quantized Nonlinear Waves}''},
\textsf{Phys.~Rev.~D12,~1643~(1975)}.
%
\bibitem{Dashen:1974ci}
R.~F.~Dashen, B.~Hasslacher and A.~Neveu,
\textit{``{Nonperturbative Methods and Extended Hadron Models in Field Theory.
  1. Semiclassical Functional Methods}''},
\textsf{Phys.~Rev.~D10,~4114~(1974)}.
%
\bibitem{Dashen:1974cj}
R.~F.~Dashen, B.~Hasslacher and A.~Neveu,
\textit{``{Nonperturbative Methods and Extended Hadron Models in Field Theory.
  2. Two-Dimensional Models and Extended Hadrons}''},
\textsf{Phys.~Rev.~D10,~4130~(1974)}.
%
\bibitem{Dashen:1974ck}
R.~F.~Dashen, B.~Hasslacher and A.~Neveu,
\textit{``{Nonperturbative Methods and Extended Hadron Models in Field Theory.
  3. Four-Dimensional Nonabelian Models}''},
\textsf{Phys.~Rev.~D10,~4138~(1974)}.
%
\bibitem{Dashen:1975hd}
R.~F.~Dashen, B.~Hasslacher and A.~Neveu,
\textit{``{The Particle Spectrum in Model Field Theories from Semiclassical
  Functional Integral Techniques}''},
\textsf{Phys.~Rev.~D11,~3424~(1975)}.
%
\bibitem{Chen:2007vs}
H.-Y.~Chen, N.~Dorey and R.~F.~L.~Matos,
\textit{``Quantum Scattering of Giant Magnons''},
\href{http://arXiv.org/abs/arXiv:0707.0668 [hep-th]}{\texttt{arXiv:0707.0668
  [hep-th]}}.
%
\bibitem{Zamolodchikov:1995xk}
A.~B.~Zamolodchikov,
\textit{``{Mass scale in the sine-Gordon model and its reductions}''},
\textsf{Int.~J.~Mod.~Phys.~A10,~1125~(1995)}.
%
%
\bibitem{Bobenko:1984}
A.~Bobenko,
\textit{``{Periodic finite-zone solutions of the sine-Gordon equation}''},
\textsf{Funkts.~Anal.~Prilozh.~18:3,~73~(1984)}.
%
\bibitem{Dorey:2006mx}
N.~Dorey and B.~Vicedo,
\textit{``{A symplectic structure for string theory on integrable
  backgrounds}''},
\textsf{JHEP~0703,~045~(2007)},
\href{http://arXiv.org/abs/hep-th/0606287}{\texttt{hep-th/0606287}}.
%
\bibitem{Mussardo:2004zn}
G.~Mussardo, V.~Riva and G.~Sotkov,
\textit{``{Semiclassical scaling functions of sine-Gordon model}''},
\textsf{Nucl.~Phys.~B699,~545~(2004)},
\href{http://arXiv.org/abs/hep-th/0405139}{\texttt{hep-th/0405139}}.
%
\bibitem{abramowitz+stegun}
M.~Abramowitz and I.~A.~Stegun,
\textit{``Handbook of mathematical functions with formulas, graphs, and
  mathematical tables''},
ninth dover printing, tenth gpo printing edition,
Dover (1964),
New York.
%
\end{thebibliography}

\end{document}